\documentclass[10pt]{article}
\usepackage{geometry}
\usepackage{xcolor}
\definecolor{graycolor}{gray}{0.9} 
\usepackage{microtype}
\usepackage{siunitx}
\usepackage{tikz}
\usepackage{setspace} 
\usepackage{amsmath} 
\usepackage[utf8]{inputenc}
\usepackage[english]{babel}
\usepackage{times}
\usepackage{array}
\usepackage{soul}
\usepackage{cite}
\usepackage{amssymb}
\usepackage[numbers,sort&compress]{natbib}
\usepackage{natbib}

\usepackage{titlesec} 
\titleformat {\section} [block] {\raggedright \fontsize{10}{10}\selectfont\bfseries} {\thesection. \space} {0pt} {}
\titlespacing {\section} {0pt} {12pt} {6pt}
\titleformat {\subsection} [block] {\raggedright \fontsize{10}{10}\selectfont\itshape} {\thesubsection .\space} {0pt} {}
\titlespacing {\subsection} {0pt} {12pt} {6pt}
\titleformat {\subsubsection} [block] {\raggedright \fontsize{10}{10}\selectfont} {\thesubsubsection .\space} {0pt} {}
\titlespacing {\subsubsection} {0pt} {12pt} {6pt}
\titleformat {\paragraph} [block] {\raggedright \fontsize{10}{10}\selectfont} {} {0pt} {}
\titlespacing {\paragraph} {0pt} {12pt} {6pt}

\usepackage{array} \newcommand{\PreserveBackslash}[1]{\let\temp=\\#1\let\\=\temp}
\newcolumntype{C}[1]{>{\PreserveBackslash\centering}m{#1}}
\newcolumntype{R}[1]{>{\PreserveBackslash\raggedleft}m{#1}}
\newcolumntype{L}[1]{>{\PreserveBackslash\raggedright}m{#1}}
\usepackage{lineno}
\usepackage{tabularx}
\usepackage{colortbl}
\usepackage{graphicx}
\usepackage{float}
\usepackage[export]{adjustbox}
\usepackage{caption}
\usepackage{ upgreek }
\usepackage{lastpage}
\usepackage{layout}
\usepackage{setspace} 
\usepackage{enumitem}
\usepackage{booktabs}
\usepackage{arydshln}
\usepackage{multirow}
\usepackage{color}
\usepackage{hyperref} 
\hypersetup{
	colorlinks=true,
	linkcolor=blue,
	filecolor=blue,
	urlcolor=black,
	citecolor=cyan,
}

\soulregister\cite7
\soulregister\ref7
\setcitestyle{open={[},close={]},citesep={,\!},numbers}
\begin{document}
\newgeometry{left=2.5cm, right=2.5cm, top=1.8cm, bottom=4cm}
\title{\bf{MeV--GeV Gamma-Ray Astrophysics in the\\Multimessenger Era}}  
	{\author{Alessandro De Angelis\\
			\small{Dipartimento di Fisica e Astronomia, Universit\`a di Padova, 35122 Padova, Italy}%; alessandro.deangelis@unipd.it 	
			}}
%		\footnotesize	\textbf{How To Cite}: {De Angelis, A. MeV--GeV Gamma-Ray Astrophysics in the Multimessenger Era. \emph{Physics and the Cosmos} \textbf{2026}, \emph{Volume}(Issue), Page Number. \href{https://doi.org/10.xxxx/xxx}{https://doi.org/10.xxxx/xxx}}}\\

\date{ }

%\begin{tabular}{lp{12cm}}  
% \small 
%  \begin{tabular}[t]{@{}l@{}} 
%  \footnotesize  Received: day month year \\
%  \footnotesize  Revised: day month year \\
%   \footnotesize Accepted: day month year \\
%  \footnotesize  Published: day month year
%  \end{tabular} &

\maketitle

\begin{abstract} Gamma-ray astrophysics probes the most extreme particle accelerators
and explosive transients in the Universe. From pioneering theoretical predictions
in the 1950s and the first space-borne detections in the 1960s, mostly exploring the sub-MeV region, the field has evolved
into a mature, multi-decade enterprise that spans nine orders of magnitude in photon
energy up to PeV energies and interfaces naturally with neutrino and gravitational-wave astronomy. Yet
the energy range from a few hundred keV to a few GeV---the “MeV gap”,
constraining progress on nucleosynthesis, positron annihilation, transient physics,
dark-matter signatures, and electromagnetic counterparts to high-energy neutrinos and
gravitational waves---remains sensitivity-limited. In this paper, we survey the scientific motivations for gamma-ray astrophysics, sketch a concise history from the first ideas to key milestones in space- and ground-
based gamma-ray astronomy, and discuss programmatic attempts to close the MeV~gap. \\
\\
  \textbf{Keywords:} Multimessenger astrophysics; Gamma-rays; MeV--GeV astronomy
\end{abstract}

\newcommand{\eam}{a new MeV telescope}

	\section{Introduction }
	Gamma rays trace non-thermal processes---shock and reconnection acceleration, magnetospheric gaps, inverse Compton scattering, hadronic interactions, and nuclear lines from freshly synthesized isotopes. They:
\begin{itemize}[topsep=3pt,parsep=0pt,itemsep=0pt,leftmargin=*,labelsep=6.2mm,align=parleft]
   \item Diagnose compact objects (pulsars, magnetars, black-hole accretion and jets), including variability on millisecond to year scales.
   \item Reveal cosmic-ray origin and feedback (supernova remnants, pulsar wind nebulae, starburst galaxies).
  \item Provide unique nucleosynthesis tracers via nuclear de-excitation lines (e.g., $^{26}$Al at 1.809~MeV; $^{60}$Fe at 1173/1333~keV) and positron annihilation at 511~keV.
    \item Characterize transients (gamma-ray bursts, magnetar flares), crucial to multi-messenger campaigns.
  \item Offer sensitive tests of fundamental physics (Lorentz-invariance, axion-like particle mixing, dark-matter annihilation/decay features).
\end{itemize}

The multi-messenger era deepens  motivations for gamma-ray astrophysics. In particular, MeV--GeV photons constrain jet baryon content in neutrino-emitting blazars; MeV lines expose $r$-process yields in neutron-star mergers that produce gravitational waves; gamma-ray detectors allow rapid, arcminute localizations that catalyze follow-ups and can in turn quickly react to alerts.

\section{A Short History of Gamma-Ray Astrophysics, Focusing on the MeV/GeV Region}

The idea of a high-energy gamma-ray sky followed shortly after modern cosmic-ray physics (see \cite{gottfried2} for a review of the instruments). %\bibitem{gottfried2}Siegert, T.,  Horan, D., \&  Kanbach, G., 2024, Telescope Concepts in Gamma-Ray Astronomy. In C. Bambi \& A. Santangelo (Eds.), Handbook of X-ray and Gamma-ray Astrophysics. Singapore: Springer Nature Singapore. arXiv:2207.02248.. 
Hayakawa \cite{Hayakawa1952} estimated in 1952 a diffuse Galactic flux from cosmic-ray interactions in the interstellar medium, and Morrison \cite{Morrison1958} in 1958 advocated a new astronomical window at $\gamma$-ray energies. Early space efforts included Explorer~11 (1961, see \cite{explorer}), followed by the key detection of $E\gtrsim 50$--70~MeV gamma rays from the Galactic plane by OSO-3 (launched 1967, see \cite{Clark1968}), confirming a bright Milky Way component and an isotropic background. SAS-2 (1972--1973, see \cite{SAS2}) and COS-B (1975--1982, see \cite{COSB}) delivered the first maps and source associations, including pulsars. 

\restoregeometry

%\bibitem{INTEGRAL} Winkler, C., et al. 2003, %\textit{A\&A}, 411, L1; ESA/CNES/GCN mission pages.
%\bibitem[Swift/BAT]{SwiftBAT} Barthelmy, S.D., et al. 2005, \textit{SSRv}, 120, 143; Swift BAT instrument pages.
%\bibitem[AGILE]{AGILE} Tavani, M., et al. 2009, \textit{A\&A}, 502, 995; ASI/INAF mission pages.
%\bibitem[Fermi]{Fermi} Atwood, W.B., et al. 2009, \textit{ApJ}, 697, 1071; NASA/GSFC mission pages.

%\bibitem[Whipple/Crab(1989)]{Weekes1989} Weekes, T.C., et al. 1989, \textit{ApJ}, 342, 379.
%\bibitem[LHAASO(2021)]{LHAASO2021} Cao, Z., et al. 2021, \textit{Nature}, 594, 33.

The ground-based very-high-energy (VHE) frontier opened decisively with the Whipple 10\,m's 1989 detection of TeV photons from the Crab Nebula \cite{Weekes1989} using the imaging atmospheric Cherenkov technique (IACT), paving the way for the new telescopes H.E.S.S., MAGIC, and VERITAS. In the ultra-high-energy regime, hybrid air-shower arrays and water Cherenkov facilities (HAWC see \cite{HAWC} and  LHAASO \cite{LHAASO2021}) recently established PeV gamma rays and a population of Galactic PeVatrons.

Today, imaging atmospheric Cherenkov arrays cover the range from about 30 GeV to some 300 TeV with arcminute-scale point-spread functions and high sensitivity. The Cherenkov Telescope Array Observatory (CTA, see \cite{CTA}) is transitioning to a world laboratory with Northern and Southern sites and staged construction, targeting about an order-of-magnitude sensitivity gain over current IACTs. While wide-field particle arrays like HAWC and LHAASO continuously survey TeV transients and extended sources beyond the TeV, in a near future they will be joined by the SWGO array \cite{SWGO} in Chile.

\begin{figure}[H]
    \centering
    \includegraphics[width=0.75\textwidth]{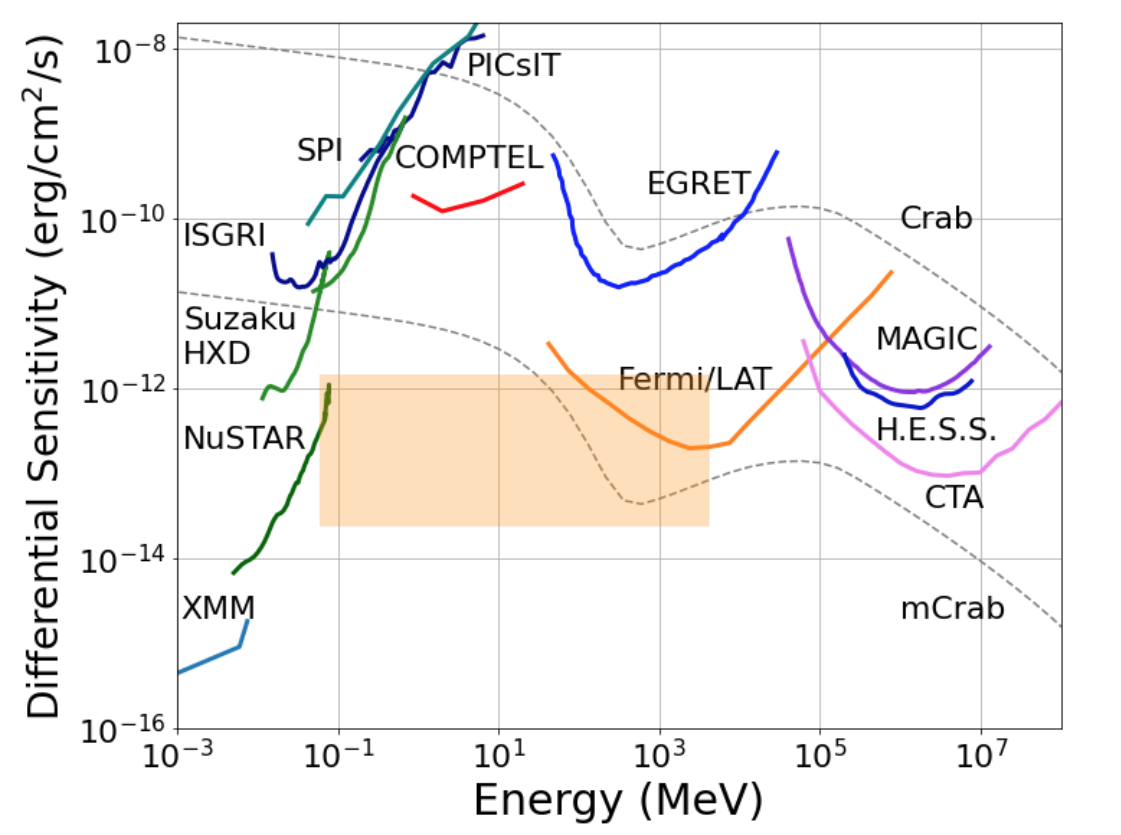}
    \caption{
    \textls[-15]{Differential sensitivity for current and past X-ray and gamma-ray missions shows the limited performance achieved in the MeV regime. 
    The reduced sensitivity in the range from 100~keV to about 3 GeV is referred to as the ``MeV gap''.
The dashed grey line representing the Crab flux is calculated from Naima~{\cite{naima}}. The figure is taken from~{\cite{gottfried1}}, in which the details of the calculations of sensitivities are explained. An orange rectangle evidences the ``MeV~gap''.}
     }
    \label{fig:sensi}
\end{figure}

The Compton Gamma Ray Observatory CGRO (1991–2000, see for example \cite{CGRO}) provided a natural link between the hard X-ray/soft gamma-ray region and the hard $\gamma$-ray region covered by IACTs. It unified the MeV--GeV band with the instruments OSSE (0.05–10 MeV), COMPTEL (0.8–30 MeV), and EGRET (20 MeV–30 GeV), plus the gamma-ray burst (GRB) dedicated BATSE. INTEGRAL (from 2002 to 2025, see \cite{INTEGRAL}) provided line spectroscopy and hard-X-ray/soft-$\gamma$  imaging; Swift/BAT (since 2004) revolutionized GRB triggering; AGILE (2007–2024, see \cite{AGILE}) and Fermi (launched 2008, see \cite{Fermi}) expanded GeV surveys and time-domain coverage. Together, CGRO’s four-in-one approach established source demographics; INTEGRAL delivered the line/continuum view and long-baseline 511 keV maps; Swift/BAT’s coded mask made rapid multi-wavelength GRB follow-up routine; AGILE pioneered fast alerts and agile silicon tracking; and Fermi’s LAT/GBM pairing remains the GeV survey workhorse and GRB trigger backbone.
Yet the soft-to-medium-energy $\gamma$-ray regime remains one of the least explored regions in multi-wavelength astrophysics. Owing to the comparatively large fluxes, photons up to—and including—the X-ray band can be measured with relatively compact instruments. But above X-ray energies and below a few tens of GeV, detection must be done from space because the atmosphere absorbs these photons, and detectors cannot be that small. The MeV--GeV band, in between these regions, is thus experimentally the most difficult. This fact becomes clear if one looks at Figure \ref{fig:sensi}, illustrating continuum sensitivities of past and current hard X-ray and  $\gamma$-ray missions and showing that performance in the MeV--GeV is orders of magnitude worse than in neighboring bands.

\vspace{-6pt}
\vspace{-16pt}

The MeV--GeV band is scientifically rich. Neutron stars, black holes, magnetars with extreme magnetic fields, and active galactic nuclei with relativistic jets naturally emit $\gamma$-rays in this energy range via bremsstrahlung, inverse Compton, and synchrotron radiation. Measurements in the MeV region constrain the thermal and non-thermal energy budgets of high-energy Galactic and extragalactic sources. This is also a promising range for indirect signatures of dark matter. Moreover, MeV energies are the natural scale for probing nuclear processes in the Galaxy and beyond: many radioactive isotopes produced in stellar and explosive nucleosynthesis emit $\gamma$-rays up to nuclear binding energies---a few MeV per nucleon---and proton-rich isotopes often undergo $\beta^+$ decay, producing positrons that annihilate and generate a characteristic 511 keV line.

Filling the “MeV gap” in sensitivity---spanning the energy region from about 100 keV to about 3 GeV---requires specialized space instrumentation. Coded-mask telescopes operate in hard X-ray region; Compton telescopes cover hundreds of keV to tens of MeV; and pair-conversion trackers extend above tens of MeV.

The shortage of sensitive MeV telescopes reflects mostly physical and technological  challenges. The photon–matter interaction cross-section is minimal in this range, requiring large detector volumes to stop and contain photons with interaction depths of about 10 g cm$^{-2}$. Without focusing optics (which is impossible in the MeV range), these large detectors face high backgrounds from atmospheric albedo and from activation within and around the instrument. While bright, short transients remain detectable, steady-source studies demand sophisticated background-rejection strategies.

\section{Key Science Drivers}
\label{sec:key_science}

MeV photon energies are crucial in different sectors of astrophysics (for a review, see \cite{eastrogam,scienceea}).

Many of the most energetic objects in the Universe have their peak emissivity at photon energies between 0.2--100~MeV 
(e.g., gamma-ray bursts, blazars, pulsars), so it is in this energy band that essential physical properties of these objects 
can be studied most directly. This energy range is also known to feature spectral characteristics associated with gamma-ray 
emission from pion decay, thus indicating hadronic acceleration. This fact makes the MeV energy region of paramount importance 
for the study of radiating, nonthermal particles and for distinguishing leptonic from hadronic processes. 

Moreover, this energy 
domain covers the crucial range of nuclear gamma-ray lines produced by radioactive decay, nuclear collision, positron annihilation, 
or neutron capture, which makes it as special for high-energy astronomy as optical spectroscopy is for phenomena related to atomic physics.

In addition, observations of relativistic jet and outflow sources (both in our Galaxy and in active galactic nuclei, AGNs) 
in the X-ray and GeV--TeV energy ranges have shown that the MeV--GeV band holds the key to understanding the transition from 
the low-energy continuum to a spectral range shaped by very poorly understood particle-acceleration processes.

A new MeV telescope with a performance similar to the one in Table~\ref{tab:perfo} would:
\begin{enumerate}[label=(\arabic*),topsep=3pt,parsep=0pt,itemsep=0pt,leftmargin=*,labelsep=3.6mm,align=parleft]
  \item determine the composition (hadronic or leptonic) of the outflows and jets---breakthrough polarimetric capability and spectroscopy providing the keys to unlocking this long-standing question;
  \item identify the physical acceleration processes in these outflows and jets (e.g., diffusive shocks, magnetic-field reconnection, plasma effects), which may lead to dramatically different particle-energy distributions;
  \item clarify the role of the magnetic field in powering ultrarelativistic jets in gamma-ray bursts, through time-resolved polarimetry and spectroscopy.
\end{enumerate}

In addition, measurements in the MeV--GeV energy band will have a major impact on multi-messenger astronomy, that will characterize science in the 2030s and beyond. 
In particular, MeV energies are expected to be the characteristic cutoffs in neutron star--neutron star (NS--NS) and black hole--neutron star (BH--NS) mergers, 
providing decisive input to the study of the energetics of these processes.

Let us analyze in more detail the main science fields  that a
MeV--GeV detector with a performance similar to that in Table \ref{tab:perfo} will open.

\begin{table}[H]
\begin{center}
\caption{\small Required  performance of a ``MeV telescope'' to achieve the core science objectives.
\label{tab:perfo}}
\newcolumntype{C}{>{\centering\arraybackslash}X} 
\begin{tabularx}{\textwidth}{cC}\toprule 
\textbf{Parameter} & \textbf{Value} \\ \midrule
Spectral range & 100~keV--3 GeV \\ \midrule
Field of view & $\geq$2.5 sr \\ \midrule
Continuum flux sensitivity  & $<$$2 \times 10^{-5}$ MeV cm$^{-2}$ s$^{-1}$ at 1 MeV (any source) \\
for $10^6$ s observation time & $<$$5 \times 10^{-5}$ MeV cm$^{-2}$ s$^{-1}$ at 10 MeV (high-latitude source) \\
($3\sigma$ confidence level) & $<$$3 \times 10^{-6}$ MeV cm$^{-2}$ s$^{-1}$ at 500 MeV (high-latitude source) \\ \midrule
Line flux sensitivity  & $<$$5 \times 10^{-6}$ ph cm$^{-2}$ s$^{-1}$ for the 511 keV line \\
for $10^6$ s observation time & $<$$5 \times 10^{-6}$ ph cm$^{-2}$ s$^{-1}$ for the 847 keV SN~Ia line \\
($3\sigma$ confidence level) & $<$$3 \times 10^{-6}$ ph cm$^{-2}$ s$^{-1}$ for the 4.44~MeV line from LECRs \\ \midrule
 & $\leq$$1.5^\circ$ at 1 MeV (FWHM of the angular resolution measure) \\
Angular resolution & $\leq$$1.5^\circ$ at 100 MeV (68\% containment radius)  \\ 
 & $\leq$$0.2^\circ$ at 1 GeV (68\% containment radius)  \\ \midrule
Polarization sensitivity & Minimum Detectable Polarization $<$$20$\% (99\% confidence level) \\
 & for a 10 mCrab source in $T_{\rm obs}=10^6$ s ($\Delta E=0.1-2$ MeV) \\ \midrule
Spectral resolution & $\Delta E / E =3$\% at 1 MeV \\
 & $\Delta E / E =30$\% at 100 MeV \\ \midrule
Time tagging accuracy & 1 $\upmu$s (at $3\sigma$) \\
\midrule
\end{tabularx}
\end{center}
\end{table}
\vspace{-24pt}

%...

\subsection{The Extreme Extragalactic Universe}
\label{sec:extreme_uni}

The distant Universe hosts sources so energetic that particles are accelerated to ultra-relativistic energies in the immediate vicinity of compact objects. Their luminosities enable detections out to high redshift, probing epochs when galaxies and black holes were rapidly assembling. A substantial fraction of this activity emerges in the MeV band, so a new $\gamma$-ray facility operating from $\sim$100~keV to a few GeV would directly image the physics near supermassive black holes, within the engines of gamma-ray bursts (GRBs), and during neutron-star (NS) mergers. By resolving how particles are energized and transported, we address a central question of high-energy astrophysics: why a tiny fraction of particles monopolize the energy budget and, through feedback, shape their environments?

GRBs peak spectrally in the MeV range. Precision polarimetry in this band would reveal the geometry and strength of magnetic fields that govern particle acceleration and radiative transfer \cite{tat18}. The same data can test Lorentz-invariance via time-of-flight and spectral distortions \cite{grbliv}, and—together with future gravitational-wave facilities—pin down the link between GRB sub-classes and compact-object mergers.

Galaxy clusters, the largest bound systems, are still forming today. Their accretion and merger shocks, as well as turbulence, energize particles over megaparsec scales. Joint MeV and radio observations break degeneracies between leptonic and hadronic scenarios, quantify the partition of energy into magnetic fields and cosmic rays, and trace feedback on the intracluster medium \cite{Brunetti14}.

\begin{figure}[H]
\centering
\includegraphics[width=0.7\textwidth]{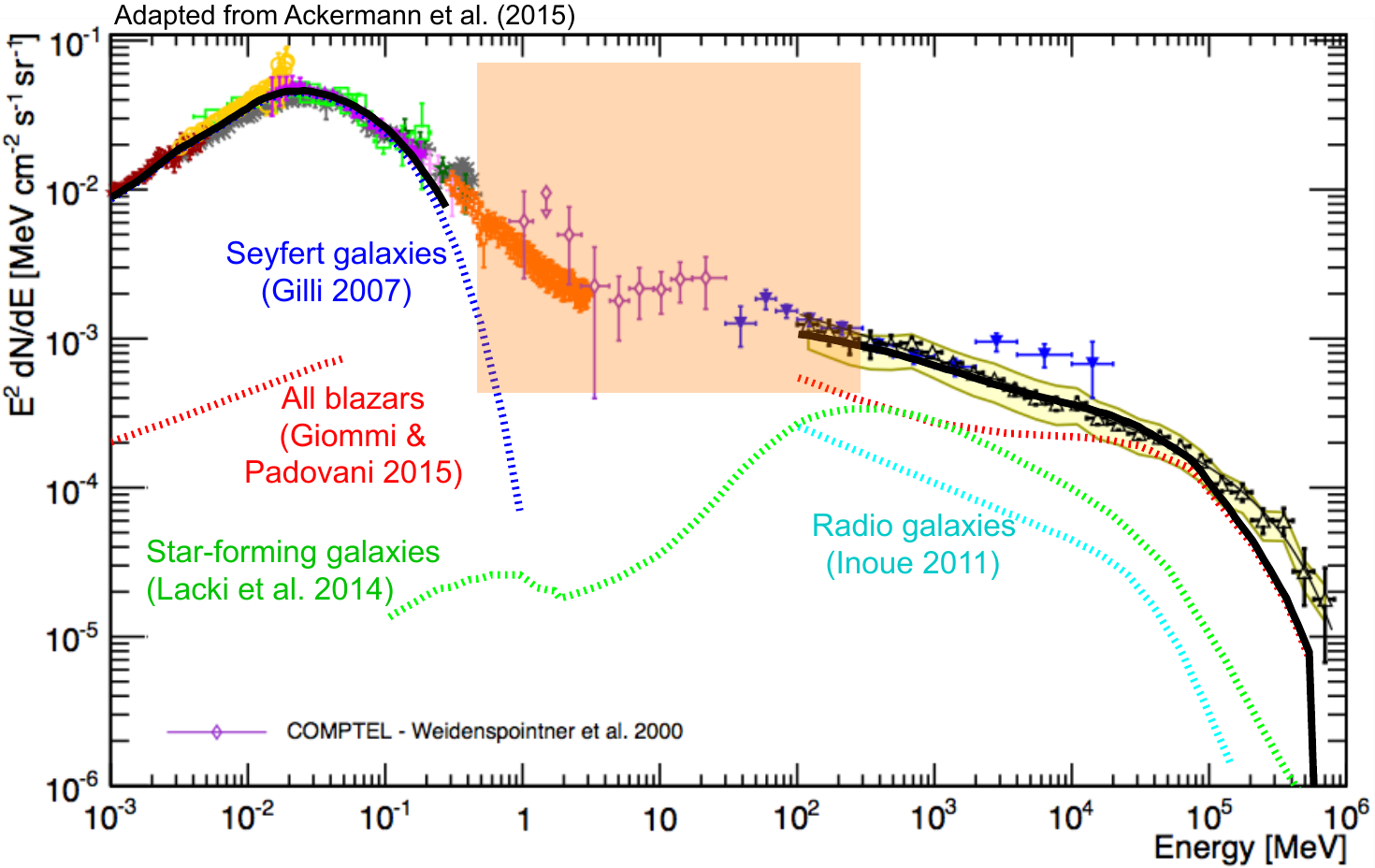}
\caption{{Measurements of the total}
 %Scilight: All mentioned references in the format of "Author+Year" in the images should be cited in the caption. Please cite all references with reference numbers and place the numbers in square brackets ("[ ]"), e.g., [1], [1-3], or [1,3].
 extragalactic $\gamma$-ray intensity from 1~keV to 820~GeV {\cite{FermiEGB}}, with schematic model 
 components \cite{gilli07,giommi15,lacki14,inoue11}; the MeV contribution of blazars remains poorly constrained. The translucent band marks the energy range where a next-generation MeV--GeV observatory could transform our knowledge. An orange rectangle evidences the ``MeV gap''.\label{fig:egb}}
\end{figure}

The MeV $\gamma$-ray background (Figure~\ref{fig:egb}) integrates emission from nucleosynthesis in distant supernovae, possible dark-matter signals, and populations of active galactic nuclei (AGN). Many AGN radiate most of their power around MeV energies, making them ideal beacons to chart the growth of supermassive black holes. Time-resolved MeV spectroscopy can locate acceleration sites, gauge the maximum electron energies, and discriminate leptonic from hadronic channels via correlated spectral changes. The MeV band is optimal: higher-energy photons suffer attenuation on the extragalactic background light, and soft $\gamma$-rays retain key diagnostics of photohadronic processes, as highlighted by the association of a $\sim$300~TeV neutrino with the flaring blazar TXS~0506+056 \cite{txsicecube,txsmulti}. Moreover, MeV measurements capture cascade emission reprocessed from absorbed very-high-energy photons, providing a census of extreme accelerators and their imprint on the intergalactic medium.

The MeV domain is also central to multi-messenger studies. The NS--NS merger GW170817, with a coincident short GRB seen by \textit{Fermi}/GBM and \textit{INTEGRAL}, demonstrated that electromagnetic cutoffs for these events lie in the MeV regime \cite{multins}. For nearby mergers, a MeV-GeV detector could detect both continuum and nuclear lines from kilonova (KN) ejecta, sharpening constraints on $r$-process nucleosynthesis \cite{pia17,multins}. Predicted line complexes in the MeV band \cite{hot16,li19} are in reach out to $\sim$10~Mpc, while current kilonova (KN) rate estimates—$\sim$200--400~Gpc$^{-3}$\,yr$^{-1}$ for GRB-associated events and $\sim$1500~Gpc$^{-3}$\,yr$^{-1}$ for GW-only detections—set realistic expectations for discoveries~\cite{del18}. If prompt MeV emission is sufficiently bright, evolving absorption edges from freshly synthesized elements may be observable \cite{ama00}.

\vspace{3pt}
\vspace{-16pt}

Finally, a telescope sensitive to the MeV--GeV region could uncover a proposed sub-class of short GRBs originating from a phase transition of a neutron star into a more compact quark star \cite{witten,perez}. Predicted signatures include ultra-brief ($\sim$0.1~s) hard spikes with $E_{\rm p}\gtrsim100$~keV, quasi-thermal components, and spectral features linked to neutron-rich ejecta, accompanied by a gravitational-wave signal distinct from binary NS/NS or NS/BH mergers. Detecting and characterizing such events would open a new window on the state of ultra-dense matter.

\subsection{Cosmic Ray Interactions}
\label{sec:cosmic_ray}

Despite a century of progress, we still lack a unified picture for the origin, acceleration, and transport of Galactic and extragalactic cosmic rays (CRs) \cite{bla13}. The problem spans plasma physics (diffusive shock acceleration and magnetic turbulence), astrophysical source modeling (injection spectra and environments), and particle diagnostics across decades in energy. Open questions range from the existence of PeVatrons in the Milky Way and EeV accelerators beyond, to the demographics and transport of sub-GeV CRs that dominate the CR energy density. A sensitive MeV--GeV $\gamma$-ray mission would address these gaps by coupling spectroscopy, imaging, and timing over the critical band where hadronic and leptonic channels can be disentangled.

Young supernova remnants (SNRs) provide benchmark laboratories. Broadband $\gamma$-ray observations across MeV--GeV energies can separate neutral-pion decay from electron bremsstrahlung and inverse-Compton components, delivering source-by-source tests of hadronic acceleration \cite{giuliani11_W44,ackermann13_W44,cardillo14_W44,jogler16,2011NatCo...2E.194M}. When combined with high-resolution radio and X-ray data, MeV--GeV spectra probe injection physics, magnetic-field structure, and the spectrum of freshly escaped CRs in interacting clouds \cite{2008ARA&A..46...89R}.

\begin{figure}[H]
\centering
\includegraphics[width=0.9\textwidth]{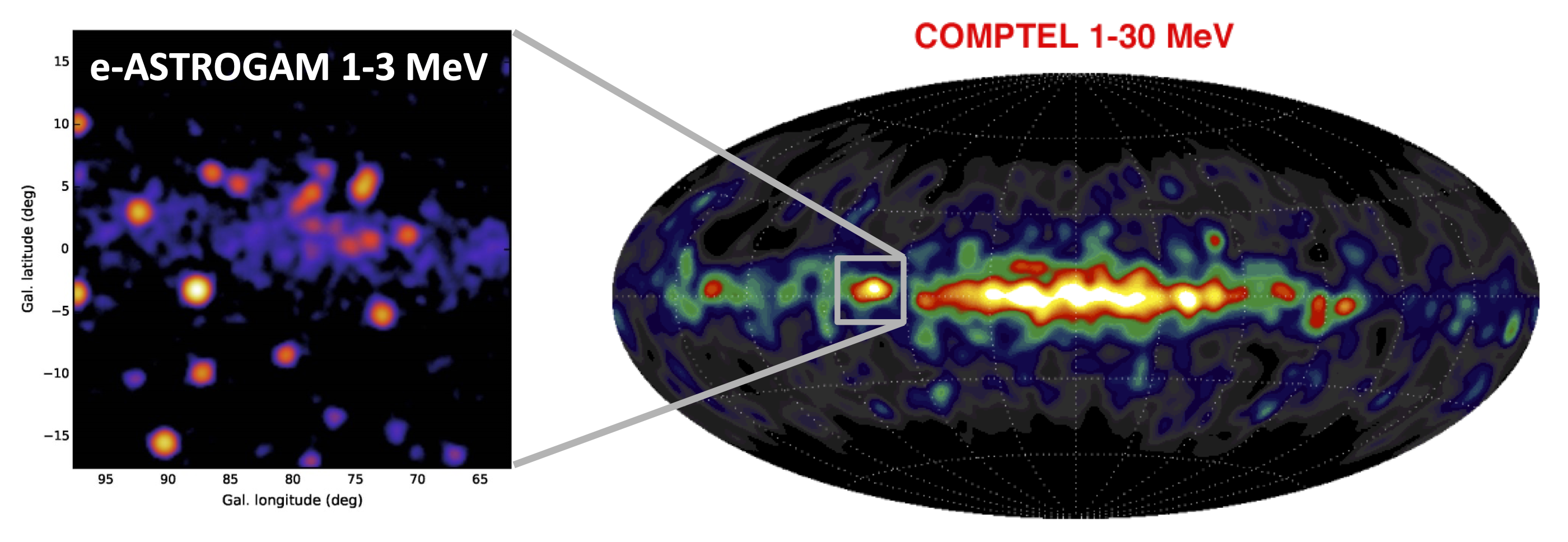}
\includegraphics[width=0.95\textwidth]{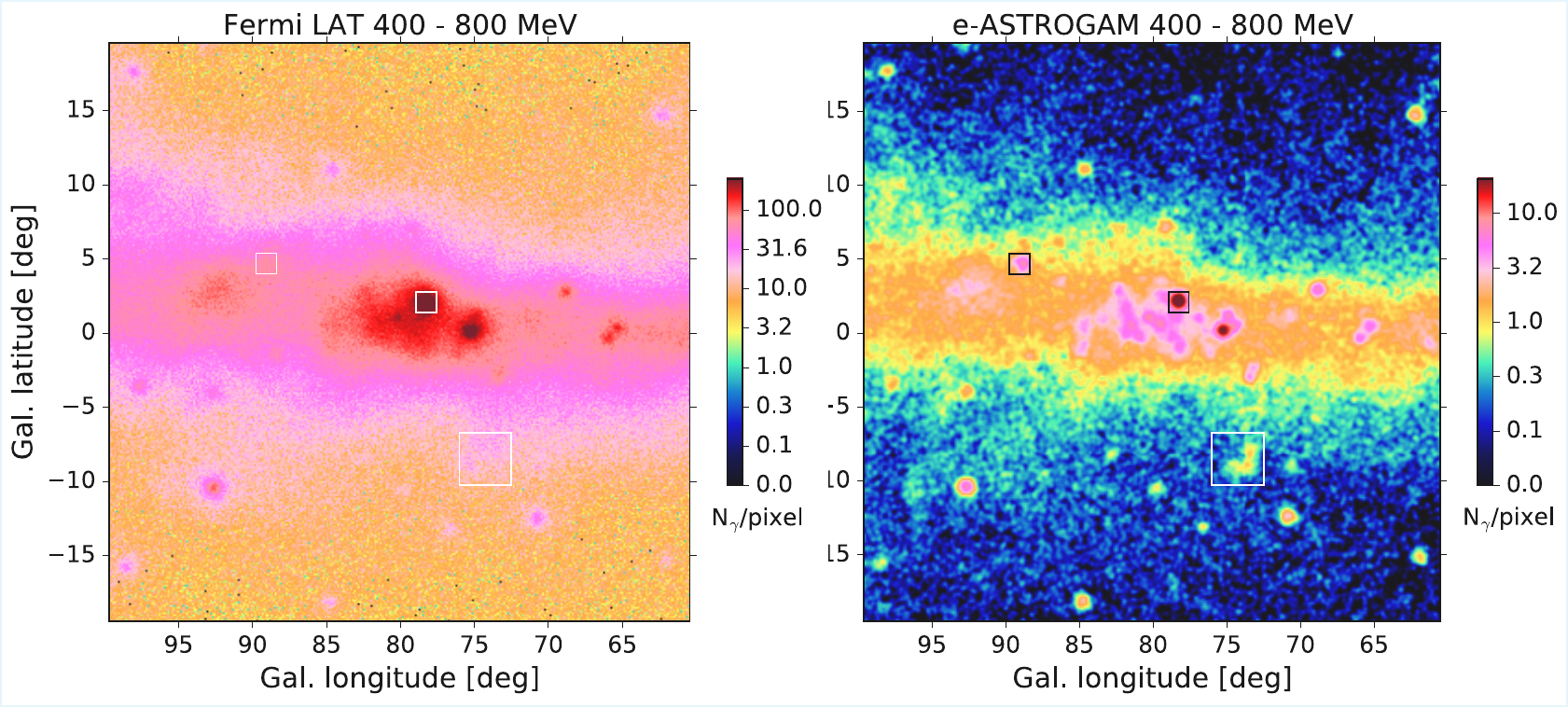}
\vspace{-2cm}
\caption{{How a next-generation} 
 MeV--GeV mission could reshape our view of the high-energy sky. Top:~\mbox{1--30~MeV} all-sky map from CGRO/COMPTEL in the 1990s ({right}) versus a simulated 1--30~MeV view of the Cygnus region ({left}). Bottom: Cygnus as seen by \textit{Fermi}-LAT after 8~yr ({left}) compared with the performance expected from a MeV--GeV detector after 1~yr between 400 and 800~MeV ({right}). The simulation is based on the performance of e-ASTROGAM.}
\label{fig:cygnus}
\label{fig:Jurgen}
\end{figure}

Beyond individual remnants, collective acceleration in superbubbles links massive-star feedback to CR production. \textit{Fermi}-LAT has firmly identified only one clear case so far—the Cygnus cocoon \cite{AckermannSB11}. Improved angular resolution below a GeV would multiply such case studies (illustrated in Figure~\ref{fig:cygnus}), enabling population work in the inner Galaxy and clarifying how CRs couple to the turbulent media of star-forming regions \cite{Bykov2014,Grenier15}. Particle acceleration in powerful stellar winds can also be tracked in extreme systems like $\eta$~Carinae, where time-domain radio-to-$\gamma$-ray monitoring constrains efficiency and environment \cite{2017arXiv170502706B}.

The \textit{Fermi} Bubbles—two giant lobes above and below the Galactic Center—remain one of the mission’s most surprising discoveries \cite{Su:2010qj,Fermi-LAT:2014sfa}. Their origin is debated between a leptonic outflow from Sgr~A* and a hadronic, starburst-driven wind. Sharper imaging and spectroscopy below 1~GeV are essential to map edges, refine morphologies, and detect the spectral curvature that distinguishes leptonic from hadronic scenarios.

Low-energy CR nuclei (sub-GeV per nucleon) carry most of the CR energy density, dominate ionization and heating in UV-shielded molecular gas, and help drive large-scale MHD outflows and Galactic winds \cite{Grenier15,Pakmor16}. MeV $\gamma$-ray spectroscopy of nuclear lines in the inner Milky Way would provide the first direct census of this population~\cite{ben13}, while continuous coverage across MeV--GeV would cleanly separate nuclei from $e^\pm$ components throughout the disk and spiral arms, constraining diffusion and source distributions \cite{Grenier15,nava17}. Finally, high-resolution GeV mapping ($\lesssim$$12'$ at 1~GeV) would deliver CR-based tracers of total gas, improving the calibration of radio and dust proxies across diverse cloud conditions \cite{Planck15}.

%\vspace{-28pt}

\subsection{Explosive Nucleosynthesis and the Galaxy’s Chemical Evolution}
\label{sec:nucleo}

Stellar explosions profoundly reshape galaxies by injecting kinetic energy and freshly forged elements into the interstellar medium, thereby steering chemical enrichment and feedback. Their “standardizable” display can even serve cosmology. Type~Ia supernovae (SNe~Ia) provided an evidence for accelerated expansion \cite{rie98,per99}; the assumption that the luminosity standardization process of Type Ia SNe remains invariant with progenitor age  has been however recently questioned (see for example \cite{son}). A deeper study of these objects is thus fundamental.  

Most transients arise either from instabilities in electron-degenerate stars (core collapse and electron-capture SNe) or from mass transfer in close binaries (SNe~Ia, classical novae). Systematic surveys now reveal a rich phenomenology \cite{hil00,hi13,woo05,jan12,bur13}.

Thermonuclear shells ignite in many of these explosions and synthesize isotopes, including radioactives. As ejecta expand and become transparent, line-bearing $\gamma$-rays escape, encoding the density/velocity structure and the distribution of freshly made nuclei. Because specific isotopes map to specific decay lines, MeV $\gamma$-ray spectroscopy is a privileged probe of explosion physics, less hindered by reprocessing than optical/IR bands \cite{isern16}. Differences among progenitors and engines thus translate into distinct line light curves, widths, and centroids that can be confronted with models.

Table~\ref{tab:radioisotopes} summarizes key lines from nucleosynthesis (see \cite{die11} for a review). Short-lived species diagnose individual events and young remnants, while long-lived isotopes produce diffuse Galactic emission from the superposition of many sources, informing stellar yields and the state of the interstellar medium \cite{die06,die16,kra15}.  

\vspace{-14pt}
\begin{table}[H]
\begin{center}
\caption{\small Star-produced radioisotopes relevant to gamma-ray line astronomy.}
\label{tab:radioisotopes}
%\vspace{2mm}
\newcolumntype{C}{>{\centering\arraybackslash}X} 
\begin{tabularx}{\textwidth}{c>{\centering\arraybackslash}m{2.6cm}>{\centering\arraybackslash}m{3cm}cC}
\toprule
{\bf Isotope} & {\bf Prod. Site $^{\rm a}$} & {\bf Decay Chain $^{\rm b}$} & {\bf Half-Life $^{\rm c}$} &
{\bf $\gamma$-ray Energy (keV)} {\bf and Intensity $^{\rm d}$} \\\midrule
$^7$Be & Nova & $^7$Be~$\stackrel{\epsilon}{\longrightarrow}$~$^7$Li*
& 53.2~d & 478~(0.10) \\\midrule
$^{56}$Ni & SNIa, CCSN &
$^{56}$Ni~$\stackrel{\epsilon}{\longrightarrow}$~$^{56}$Co*
& 6.075~d & 158~(0.99), 812~(0.86) \\
& & $^{56}$Co~$\stackrel{\epsilon(0.81)}{\longrightarrow}$~$^{56}$Fe*
& 77.2~d & {847}~(1), {1238}~(0.66) \\\midrule
$^{57}$Ni & SNIa, CCSN &
$^{57}$Ni~$\stackrel{\epsilon(0.56)}{\longrightarrow}$~$^{57}$Co*
& 1.48~d & 1378~(0.82) \\
& & $^{57}$Co~$\stackrel{\epsilon}{\longrightarrow}$~$^{57}$Fe*
& 272~d & {122}~(0.86), {136}~(0.11) \\\midrule
$^{22}$Na & Nova &
$^{22}$Na~$\stackrel{\beta^+(0.90)}{\longrightarrow}$~$^{22}$Ne*
& 2.60~y & {1275}~(1) \\\midrule
$^{44}$Ti & CCSN, SNIa &
$^{44}$Ti~$\stackrel{\epsilon}{\longrightarrow}$~$^{44}$Sc*
& 60.0~y & {68}~(0.93), {78}~(0.96) \\
& & $^{44}$Sc~$\stackrel{\beta^+(0.94)}{\longrightarrow}$~$^{44}$Ca*
& 3.97~h & {1157}~(1) \\\midrule
$^{26}$Al & CCSN, WR &
$^{26}$Al~$\stackrel{\beta^+(0.82)}{\longrightarrow}$~$^{26}$Mg*
& 7.2$\cdot$10$^5$~y & {1809}~(1) \\
 & AGB, Nova & & & \\\midrule
$^{60}$Fe & CCSN &
$^{60}$Fe~$\stackrel{\beta^-}{\longrightarrow}$~$^{60}$Co*
& 2.6$\cdot$10$^6$~y & 59~(0.02) \\
& & $^{60}$Co~$\stackrel{\beta^-}{\longrightarrow}$~$^{60}$Ni*
& 5.27~y & 1173~(1), 1332~(1) \\\bottomrule
\end{tabularx}
\begin{minipage}{\linewidth}
{ \vspace{3pt}\footnotesize
{$^{\rm a}$ Sites which are believed to produce observable
gamma-ray line emission. Nova: classical nova; SNIa: thermonuclear
SN (type Ia); CCSN: core-collapse SN; WR: Wolf-Rayet star;
AGB: asymptotic giant branch star.}
{$^{\rm b}$ $\epsilon$: orbital electron capture. When an
isotope decays by a combination of $\epsilon$ and $\beta^+$ emission, only 
the most probable decay mode is given, with the corresponding fraction in 
parenthesis.}
{$^{\rm c}$ Half-lives of the isotopes decaying by 
$\epsilon$ are for the neutral atoms.}
{$^{\rm d}$ The values in brackets correspond to the number of photons emitted in the gamma-ray
line per radioactive decay.}
}
\end{minipage}\vspace{-16pt}
\end{center}
\end{table}

%\begin{figure}[tb]
%\begin{center}
%\includegraphics[width=0.6\textwidth]{figures/Fig_DiffuseMilkyWay_RoD.pdf}
%\caption{\small Multi-band view of Galactic diffuse emission. A telescope sensitive to the MeV--GeV region would bridge starlight and cosmic-ray tracers, and refine maps of positron annihilation and $^{26}$Al radioactivity. (Composite by R.~Diehl; data from WMAP, 2MASS, \textit{INTEGRAL}, CGRO, and \textit{Fermi}; \cite{3FGL,Bennett2003,Skrutskie2006,sie16,Diehl95}).}
%\label{fig:diffuse}
%\end{center}
%\end{figure}

\begin{figure}[H]
\centering
\includegraphics[width=0.7\textwidth]{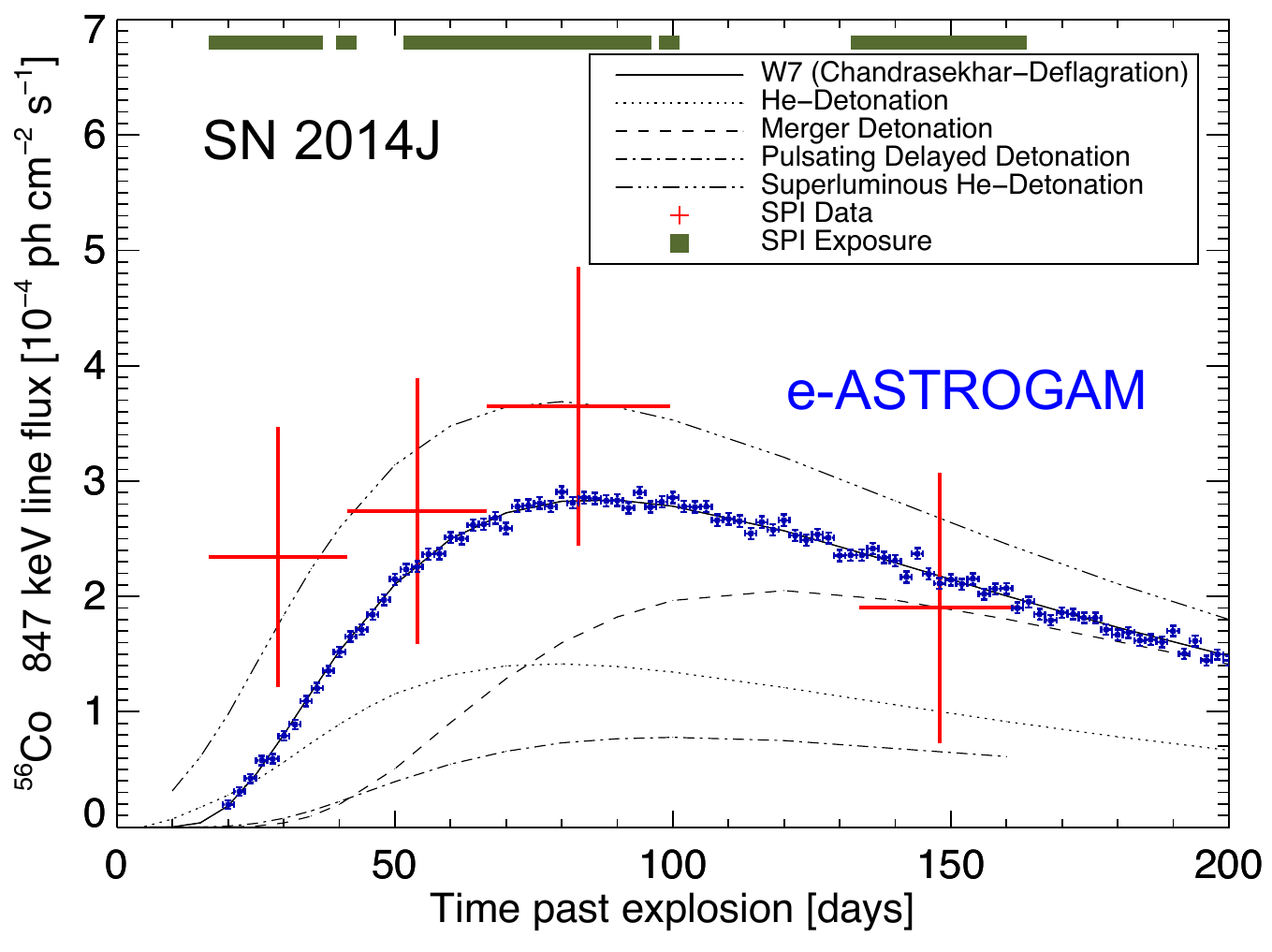}
\caption{\small Evolution of the 847~keV line from $^{56}$Co decay in SN~2014J. \textit{INTEGRAL} measurements (red; adapted from Figure~4 of {\cite{diehl15}}) are compared with SNe~Ia models {\cite{the14}}. A simulated response for a next-generation MeV mission~{\cite{eastrogam}} (blue marks), assuming W7-like evolution {\cite{nomoto84}}, illustrates how improved sensitivity can tightly constrain progenitor mass, $^{56}$Ni yield, and ejecta kinematics.}
\label{fig:SNIa}
\end{figure}

It is useful to separate \emph{planned} from \emph{opportunity} observations. Planned observations can be scheduled with reasonable certainty:
\begin{enumerate}[topsep=3pt,parsep=0pt,itemsep=0pt,leftmargin=*,labelsep=5mm,align=parleft]
\item {{SNe~Ia $^{56}$Ni/$^{56}$Co yields.}} %Scilight: Please confirm if the bold is unnecessary and can be removed. The following highlights are the same.
 Measuring the ejected $^{56}$Ni/$^{56}$Co calibrates the Phillips relation \cite{phi93} and iron production. With the sensitivity of a next-generation MeV mission \cite{eastrogam}, roughly a dozen SNe~Ia within $\lesssim35$~Mpc over three years are expected. Observations near 50--100~days post-explosion—once the subtype and luminosity are known \cite{chur14,chur15,diehl15}—would yield percent-level constraints (Figure~\ref{fig:SNIa}). 
\item {{Asymmetries from $^{44}$Ti/$^{44}$Sc.}} Mapping the 68/78/1157~keV lines probes clumping and internal asymmetries in young core-collapse remnants \cite{mah88,tue90,gre14,gre17}, including SN~1987A.
\item {{Galactic diffuse lines and positrons.}} High-fidelity maps of 511~keV annihilation and of $^{26}$Al/$^{60}$Fe emission will illuminate massive-star, SN, and nova nucleosynthesis and Galactic dynamics, with improved correlation to gas/dust tracers \cite{joh72,lev78,kno05,wei08,sie16,Diehl06,Wang07,Wang09,mar09,Diehl13,kra15}.
\end{enumerate}

Targets of Opportunity (ToOs) are unpredictable but exceptionally informative:
\begin{enumerate}[topsep=3pt,parsep=0pt,itemsep=0pt,leftmargin=*,labelsep=5mm,align=parleft]
\item {{Novae.}} The 1275~keV line from $^{22}$Na should be detectable to distances allowing $\sim$annual events, whereas the 478~keV $^{7}$Be line requires closer novae; yields and ejecta physics can be constrained on a case-by-case basis~\cite{CH74,Gom04,Her08}.
\item {{SNe~Ia and core-collapse SNe.}} Early (pre-maximum) $\gamma$-ray emission in SNe~Ia \cite{diehl14,isern16} and the $^{56}$Ni mass in core-collapse SNe \cite{mat88} are both within reach out to $\sim$10~Mpc, enabling subtype comparisons and tests of explosion channels.
\end{enumerate}

While ToOs cannot be guaranteed in advance, the step-change in sensitivity thanks to a new telescope with the appropriate performance in the MeV--GeV region ensures that a statistically meaningful sample of explosive transients will be captured and exploited effectively.

\vspace{-26pt}

\subsection{Observatory Science in the MeV Domain}

Since the MeV domain is largely unexplored, the observatory science could be particularly interesting. We summarize here some of the topics, referring the reader  to \cite{gottfried1,gottfried2} for a more complete treatment. Observatory science is particularly relevant in a multiwavelength, multimessenger context.

\subsubsection{Physics of Compact Objects}
\label{sec:comp_obj}

Neutron stars (NSs) and black holes (BHs) are extreme laboratories where strong gravity and ultra-intense magnetic fields shape the high-energy sky. They appear as rotators, bursters, accretors, wind interactors, and mergers, in both isolated and binary systems. Untangling the evolutionary links among NS sub-classes—especially the connection between rotation-powered pulsars and magnetars—requires measurements that track spectra, timing, and polarization across the poorly explored MeV window.

Magnetars offer the clearest case for MeV diagnostics: their flares and hard, non-thermal tails beyond $\sim$100\,keV point to twisted magnetospheres where magnetic reconnection, crustal failures, photon splitting, and pair cascades compete. Upper limits suggest spectral breaks in the MeV range, yet the geometry and current-carrying bundles remain uncertain. A telescope sensitive in the MeV--GeV region with phase-resolved spectroscopy and polarimetry can localize emission zones, measure the expected cutoffs, and test magnetospheric dissipation models~\cite{KaspiBeloborodov2017}.

For rotation-powered pulsars, \textit{Fermi}-LAT revolutionized the field above 100\,MeV (from 7 to $\sim$200 detections), but the soft $\gamma$-ray band holds only a few dozen sources and just a handful with pulsed emission at 1–10\,MeV. Several “MeV pulsars’’ peak in this band, implying that the next leap in population and physical understanding will come from MeV-sensitive surveys. Polarization at MeV energies can break degeneracies in inclination and viewing geometry, reveal field topology, and pinpoint emission regions in the magnetosphere \cite{Abdo2013,KuiperHermsen2015}.

Compact binaries add further leverage. Intra-binary shocks in black-widow/redback systems and in $\gamma$-ray binaries (young pulsar + Be star) accelerate pairs to multi-TeV energies. The MeV band anchors the turnover connecting X-rays to GeV photons, discriminating synchrotron from inverse-Compton scenarios and constraining particle acceleration, magnetization, and target-photon fields. A new generation Mev-Gev detector would thus provide decisive tests of competing models \cite{Dubus2013}.

Accreting NS/BH X-ray binaries may launch jets and produce hard $\gamma$-ray components; a narrow 2.2\,MeV neutron-capture line—from the inner disc or an NS atmosphere—would be a landmark detection, simultaneously probing accretion physics and the gravitational redshift at the stellar surface. Transitional millisecond pulsars that flip between rotation-powered and accretion-powered states are natural laboratories for magnetosphere–disc coupling; continuous coverage from 0.1–100\,MeV would track spectral gaps through state changes and bound acceleration limits in pulsar-wind shocks \cite{Papitto2013}.

Finally, globular clusters host rich millisecond-pulsar populations and show diffuse X-ray/TeV emission of uncertain origin. MeV imaging and spectroscopy can map any extended component, separate magnetospheric emission from pair outflows, and complement \textit{Fermi} detections of cluster $\gamma$-rays. By combining timing, spectra, and polarization, an appropriate Mev-Gev telescope would close the MeV gap and deliver a step-change in compact-object astrophysics \cite{Abdo2009}.

\subsubsection{Solar- and Earth-Science}
\label{sec:solar}

Cosmic rays (CRs) interacting with matter and radiation near the Sun, Moon, and Earth, together with particle acceleration in magnetized plasmas and magnetic reconnection, provide local laboratories to test high-energy processes that also power phenomena at galactic and extragalactic scales \cite{DiMatteo98}. A telescope with the appropriate sensitivity in the MeV--GeV region  would unify these studies by coupling broad energy coverage with timing, imaging, spectroscopy, and polarization.

Solar flares—often linked to coronal mass ejections—release magnetic energy impulsively, accelerating electrons and ions. The resulting emission spans non-thermal bremsstrahlung (hard X-rays) and nuclear/mesonic channels that yield de-excitation lines (1–10 MeV) and a $\gtrsim100$ MeV continuum (from $\pi^0$ decay). Some events also show hours-long, high-energy emission after the impulsive phase. Measurements from $\sim$100 keV to a few GeV can track the temporal evolution of each component, separate electron bremsstrahlung from pion decay, detect ion-produced nuclear lines (probing composition and interaction sites), and measure polarization to test acceleration anisotropies and reconnection-driven geometry \cite{DiMatteo98,Abdo2011}.

The Moon is a bright high-energy $\gamma$-ray source, produced when Galactic CR nuclei strike the regolith and initiate hadronic cascades. Its flux and spectrum encode the incident CR composition and the level of solar modulation. Extending measurements down to the sub-GeV/MeV domain would enable searches for narrow $\gamma$-ray lines from nuclear de-excitations in lunar rock, while continuous monitoring would turn the lunar signal into a stable CR/solar-cycle gauge. Improved low-energy spectroscopy and imaging with a Mev-GeV telescope with appropriate performance would refine the lunar spectrum and its variability \cite{Abdo2012}.

Terrestrial $\gamma$-ray flashes (TGFs) arise in the upper atmosphere and are tightly correlated with lightning, consistent with bremsstrahlung from electrons accelerated to relativistic energies in thunderstorm electric fields. Low-Earth-orbit observations over tropical storm belts maximize detection rates and enable studies of the highest-energy TGFs ($\gtrsim$40 MeV). With fast timing and broad energy reach, measurements from $\sim$100 keV to a few GeV can characterize TGF spectra and cutoff energies, constrain acceleration and feedback models, and connect atmospheric electricity to high-energy particle physics \cite{Dwyer2012}.

\subsubsection{Fundamental Physics}
\label{sec:fund_phys}

Gamma-ray astrophysics offers a unique arena to test the most basic symmetries and ingredients of nature across cosmological distances,  and in environments unreachable on Earth. Gamma rays are precision messengers: they propagate directly from their production sites, encode spectral, temporal, and polarization information, and provide leverage on questions ranging from Lorentz symmetry and  tests to the particle identity of dark matter (DM). While multiple, independent observations strongly support the presence of non-baryonic DM in the Universe, its microphysical properties remain an open problem \cite{Ade:2015xua,bergstrom12}.
A long-standing, well-motivated class of candidates is that of Weakly Interacting Massive Particles (WIMPs), with masses and couplings near the electroweak scale. Their present-day abundance is naturally explained by thermal freeze-out in the early Universe \cite{Jungman:1995df}. 

That broader landscape naturally includes models with masses at or below the GeV scale and scenarios with hidden sectors or light mediators, where the phenomenology is shifted toward lower photon energies and distinctive spectral features \cite{Feng:2008ya}. Here the MeV band becomes pivotal. A next-generation MeV gamma-ray observatory, coupling high sensitivity with good energy and angular resolution and (ideally) polarimetric capability, can open discovery space in several ways. First, it can probe final-state radiation, internal bremsstrahlung, or radiative decay signatures that peak in the $\sim$MeV–GeV domain for sub-GeV candidates. Second, it can measure spectral cutoffs, bumps, and narrow features that are either washed out at higher energies or buried under backgrounds at lower energies. Third, polarization measurements—where feasible—can help disentangle emission mechanisms, tightening indirect limits or increasing the robustness of a potential detection.

Target selection remains central to maximizing discovery potential. Regions of high DM density such as the Galactic Center (GC) deliver the largest expected flux but are complicated by bright and structured astrophysical emission. Dwarf spheroidal galaxies offer exceptionally clean conditions: their stellar kinematics enable independent inferences on the DM distribution, and their intrinsic gamma-ray emission is minimal, making them gold-standard targets for stacked analyses. Galaxy clusters become competitive in scenarios with substantial substructure boosting. Diffuse backgrounds—Galactic and extragalactic—encode additional information accessible through anisotropy studies and cross-correlation with external tracers (e.g., galaxy catalogs, weak lensing maps), providing complementary constraints to pointed observations \cite{Conrad:2015bsa,Gaskins:2016cha}. Across all these targets, a MeV instrument would bridge the current gap between hard X-ray and GeV observations, improve control of systematics through broader spectral coverage, and sharpen synergy with neutrino, gravitational-wave, and direct-detection programs.

In summary, precision gamma-ray measurements in the MeV range are a critical ingredient of the global strategy to uncover the particle nature of dark matter. They enhance sensitivity to light-mass scenarios, add independent handles on astrophysical systematics, and provide unique spectral and polarization diagnostics that complement both GeV–TeV gamma-ray instruments and non-photonic probes \cite{Ade:2015xua,bergstrom12,Jungman:1995df,Liu:2017drf,Aprile:2017iyp,Conrad:2015bsa,Gaskins:2016cha,Feng:2008ya}.

%\newpage

Another target for new physics appropriate for a MeV--GeV detector is that of axion-like particles (ALPs), very light, neutral pseudo-scalars predicted in many extensions of the Standard Model (notably string-inspired scenarios). They couple to two photons with strength $g_{a\gamma}$, while---unlike QCD axions---the axion mass $m_a$ and $g_{a\gamma}$ are  unrelated~\cite{JaeckelRingwald2010}. 
Non-observation of solar ALPs in the CAST helioscope sets a benchmark limit $g_{a\gamma}\!<\!0.66\times10^{-10}\,{\rm GeV}^{-1}$ for $m_a\!<\!0.02\,{\rm eV}$ \cite{CAST2017}.

In an external magnetic field $\mathbf{B}$, photons oscillate into ALPs and back, modifying $\gamma$-ray spectra and polarization. For a monochromatic beam of energy $E$ propagating a distance $y$ in a homogeneous region, the conversion probability reads \cite{DeAngelis2011}
\begin{equation}
P_{\gamma\to a}(E;y)=\left(\frac{g_{a\gamma}B}{\Delta_{\rm osc}}\right)^{\!2}\sin^2\!\left(\frac{\Delta_{\rm osc}y}{2}\right),\qquad
\Delta_{\rm osc}=\Bigg[\frac{(m_a^2-\omega_{\rm pl}^2)^2}{4E^2}+g_{a\gamma}^2B^2\Bigg]^{1/2},
\end{equation}
with plasma frequency $\omega_{\rm pl}$. Defining the characteristic energy
\begin{equation}
E_\ast\equiv\frac{|m_a^2-\omega_{\rm pl}^2|}{2g_{a\gamma}B}\,,
\end{equation}
one finds three regimes: no mixing for $E\ll E_\ast$, rapid energy-dependent oscillations for $E\sim E_\ast$ (weak mixing), and maximal, energy-independent mixing for $E\gg E_\ast$ (strong mixing). Extragalactic fields are often modeled as domains (size $L_{\rm dom}\!\sim\!1$--$10$ Mpc, $B\!\sim\!0.1$--$1$ nG) with random $\mathbf{B}$ orientation, which enhances the spectral wiggles around $E_\ast$; over many domains the two photon polarizations and the ALP state approach equipartition, dimming the average photon flux by a factor $\simeq2/3$ \cite{DeAngelis2011}. The same coupling induces birefringence/dichroism, so MeV polarimetry delivers an additional, independent handle on ALPs.

The MeV--GeV band  is therefore a prime hunting ground: it can capture both the spectral distortions near $E_\ast$ and polarization signatures, and it is pivotal for transient multi-messenger sources. In particular, ALPs produced via the Primakoff process in core-collapse supernovae can reconvert into $\gamma$-rays in the Galactic magnetic field; the resulting prompt $\gamma$-ray burst would be coincident with the neutrino burst and predominantly below $\sim$100 MeV, placing it squarely in the MeV domain \cite{Meyer2017}.

Finally, a possible target of a MeV--GeV detector can be  the detection of primordial black holes (PBHs): see for example \cite{scienceea}. Depending on formation time, PBH masses can span from sub-planetary scales to millions of solar masses, with different astrophysical and cosmological imprints. In the absence of significant accretion, Hawking radiation drives a secular mass loss with temperature $T_{\rm BH}\propto 1/M_{\rm BH}$ and a lifetime $\tau\propto M_{\rm BH}^{3}$, implying that PBHs lighter than $\sim$$10^{14}\,$g have already evaporated, whereas those with $M_{\rm BH}\gtrsim10^{15}\,$g can survive to the present and potentially contribute to the dark matter (DM) inventory. Photon emission from Hawking radiation comprises a primary component directly from the black hole and a secondary one from hadronization and decays of emitted quarks, gluons, and gauge bosons; the latter produces a robust spectral bump near $E_\gamma\!\sim\!70$\,MeV due to $\pi^0\!\to\!\gamma\gamma$, weakly dependent on $T_{\rm BH}$. This anchors the MeV–sub-GeV band as a privileged discovery space: PBHs with masses $\sim$$10^{15\text{--}17}\,$g are expected to peak in the $\sim$1--30\,MeV output, so their cumulative emission would contribute to the diffuse extragalactic $\gamma$-ray background (EGB), and improved spectroscopy in this window can tighten or reveal their signal.

\section{How to Detect Gamma Rays: Interactions at Energies Around the MeV}\label{sec:physics}

At high photon energies, like $\gamma$-ray energies,  interactions are rather particle-like than wave-like and can be exploited to measure the photon’s energy via its partial or total energy deposit in the detector. 

For $\gamma$-ray photons, the three dominant processes of interaction with matter, that can thus be exploited in detectors, are \cite{noibook}:
\begin{enumerate}[topsep=3pt,parsep=0pt,itemsep=0pt,leftmargin=*,labelsep=5mm,align=parleft]
\item Photoelectric absorption. 
\item Compton scattering.
\item Pair production.
\end{enumerate}

The rate of survival of a beam  is usually expressed via the interaction length $\ell = 1/n\sigma$, where $\sigma$ is the cross section and $n$ is the number density of targets:
\begin{equation}
   I=I_0\,e^{-x/\ell} \, ,
\end{equation}
where $I_0$ is the initial intensity  and $I$ is the intensity after traversing a path $x$.
Total interaction probability scales with the number of targets, i.e., for the same material, with matter density; it is thus convenient to refer to scaled cross sections $\sigma$ in barn per atom or in barn per unit density to estimate beam transmission.

\begin{figure}[H]
    \centering
    \includegraphics[width=0.5\textwidth]{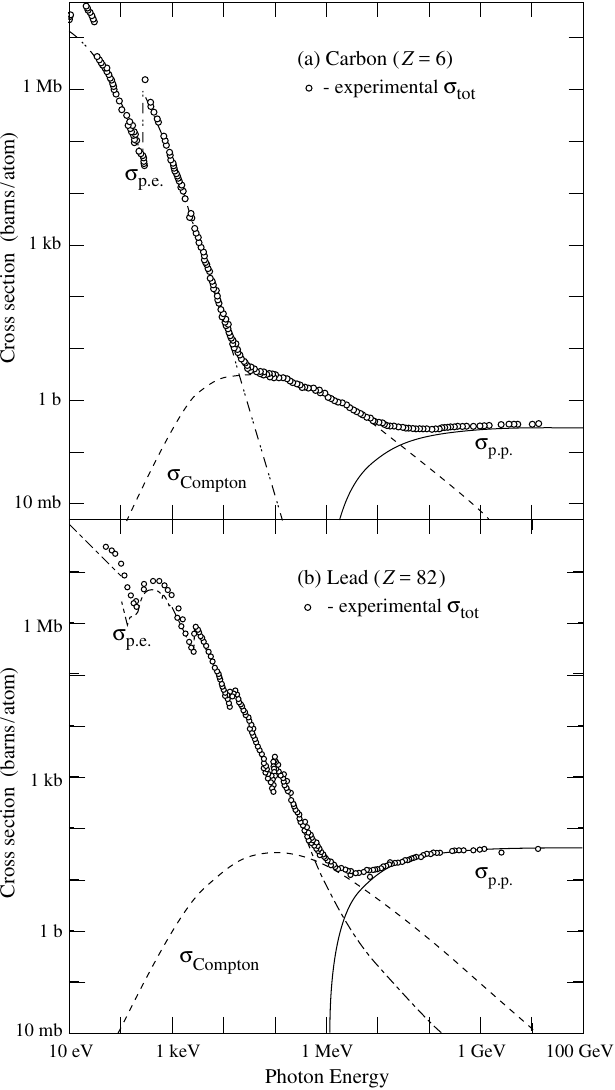}
    \caption{
    Photon total interaction cross sections as a function of energy in carbon and lead, showing the
contributions of different dominant processes: 
$\sigma_{\rm{p.e.}}$ = Photoelectric effect (electron ejection, photon absorption); $\sigma_{\rm{Compton}}$ = Incoherent scattering (Compton scattering off an electron);
$\sigma_{p.p.}$ = Pair production. The open circles show experimental points for the total cross section.
    The figure is taken from \cite{pdg}.}
    \label{fig:xsec}
\end{figure}

 Figure \ref{fig:xsec} shows the cross section $\sigma$ in barn per atom 
 for a low-$Z$ and a high-$Z$ material.
 While the qualitative shapes of photoelectric, Compton, and pair-production components are similar across materials, details shift operating regimes and guide detector choices. For example, plastic scintillator exhibits a broad Compton-dominant band, making it a preferred scatterer in classic Compton detector designs.

 \vspace{6pt}
\vspace{-26pt}

\subsection{Photoelectric Effect\index{Photoelectric effect}}

The photoelectric effect is the ejection of an electron from a
material that has just absorbed a photon. The ejected electron is
called a  {\em photoelectron}.

Due to photoelectric effect, a photon of
angular frequency $\hbar\omega > V$ can eject from a material an electron
which pops up with a kinetic energy $\hbar \omega - V$, where $V$ is
the minimum gap of energy of electrons trapped in the material ($V$ is
frequently called the {\em work function} of the material itself).

No simple relationship between the attenuation of the incident
electromagnetic wave and the photon energy $E$ can be derived, since
the process is characterized by the interaction with the (quantized)
orbitals. The plot of the attenuation coefficient  (the distance per
unit density  at which intensity is reduced by a factor $1/e$) as a
function of the photon energy displays sharp peaks at the binding
energies of the different orbital shells and depends strongly on the
atomic number. Neglecting these effects, a reasonable approximation for the cross section
$\sigma$ is
\[ \sigma \propto \frac{Z^\nu}{E^3} \, ,\]
with the exponent $\nu$ varying between 4 and 5 depending on the
energy.
The cross section rapidly decreases with energy above the typical
electron binding energies (Figure\,\ref{fig:xsec}).

The photoelectric effect can be used for detecting photons below the
MeV; a  photosensor (see later) sensitive to such energies can
``read'' the signal generated by a photoelectron, possibly amplified
by an avalanche process.

\subsection{Compton Scattering\index{Compton!scattering}}

Compton scattering is the collision between a photon and an
electron. Let $E$ be the energy of the primary photon (corresponding
to a wavelength $\lambda$) and suppose that the electron is initially
free and at rest. After the collision, the photon is scattered at an
angle $\theta$ and comes out with a reduced energy $E'$,
corresponding to a wavelength $\lambda'$; the electron acquires an
energy $E-E'$. The conservation laws of energy and momentum yield
the following relation  (Compton formula):
\[ \lambda' - \lambda = \lambda_C (1- \cos \theta) \longrightarrow E' = \frac{E}{1+ \frac{E}{m_ec^2}(1-\cos \theta)} \]
where $\theta$ is the scattering angle of the emitted photon;
$\lambda_C = h/m_e c \simeq 2.4$ pm is the Compton wavelength of the~electron.

It should be noted that, in the case when the target electron is not at rest,
the energy of the scattered photon can be larger than the energy of
the incoming one. This regime is called {\em inverse
Compton,} and\index{Inverse Compton} it has great importance in the
emission of high-energy photons by astrophysical sources: in
practice, thanks to inverse Compton, photons can be ``accelerated.''

The differential cross section for Compton scattering was calculated
by Klein and Nishina around 1930. If the photon energy is much below
$m_e c^2$ (so the scattered electrons are non-relativistic) then the
total cross section is given by the Thomson cross section. This is
known as the Thomson limit. The cross section for $E \ll m_e c^2$
(Thomson regime) is about
\begin{equation}
\sigma_T \simeq \frac{8\pi \alpha^2}{3 m^2_e} = \frac{8\pi r_e^2}{3} \, ,
\end{equation}
where $r_e = (e^2/4\pi\epsilon_0)/(m_e c^2) \simeq$ 0.003 pm is the
classical radius of the electron. For photon energies well above the electron rest mass ($x \equiv E/(m_ec^2)\gg 1$), the total Klein--Nishina
cross-section decreases approximately as
\begin{equation}
  \sigma_{KN}(E)\;\simeq\;\frac{3}{8}\,\sigma_T\,\frac{\ln(2x)+\tfrac{1}{2}}{x},
  \qquad x \equiv \frac{E}{m_ec^2},
\end{equation}
illustrating the rapid fall-off of the Compton scattering probability at high energy.

As in the case of the photoelectric effect, the ejected electron can
be detected (possibly after multiplication) by an appropriate
sensor.

\subsection{Pair Production\index{Pair production}}

Pair production is the most important interaction process for a
photon above an energy of a few tens of  MeV. In the electric field
in the neighborhood of a nucleus, a high-energy photon has a non-negligible probability of transforming itself into a negative and a
positive electron---the process being kinematically forbidden
unless an external field, regardless of how little, is present.

Energy conservation yields the following relation between the energy
$E$ of the primary photon and the total energies $U$ and $U'$ of the
electrons:
\[ E = U + U'. \]

With reasonable approximation,  for  1 TeV$\,> E > 100$\,MeV the
fraction of energy $u = U/E$ taken by the secondary
electron/positron is  uniformly distributed between 0 and 1
(becoming peaked at the extremes as the energy increases to values
above 1\,PeV).

The cross section grows quickly from the kinematic threshold of
about 1 MeV to its asymptotic value reached at some 100 MeV:
\[ \sigma \simeq \frac{7}{9}\frac{1}{n_a X_0} \, ,\]
where $n_a$ is the density of atoms per unit volume, in such
a way that the interaction length is
\[ \ell \simeq \frac{9}{7}X_0 \]
($X_0$ is called the {\em radiation length} of the material).

The angle of emission for the particles in the pair is typically
${\sim}0.8 \,{\rm{MeV}}/E$.

\subsection{Comparison Between Different Processes for Photons}

The total Compton scattering probability decreases rapidly when the
photon energy increases. Conversely, the total pair production
probability is a slowly increasing function of energy. At large
energies, most photons are thus absorbed by pair production, while
photon absorption by the Compton effect dominates at low energies
(being the photoelectric effect characteristic of even smaller
energies). 

As a matter of fact, above about 20\,MeV the dominant process is pair
production, and the interaction length of a photon is, to an extremely
good approximation, equal to $9X_0/7$.

\section{Programmatic Attempts for a New MeV Detector}

A sensitive telescope for the MeV gap must exploit the dominant interaction in the MeV band: Compton scattering. In Compton events, an incident $\gamma$  ray transfers part of its energy to a (possibly bound) electron, which recoils (and might possibly be tracked) while the photon is deflected with reduced energy. The scattered photon then interacts again, potentially multiple times, until its energy is fully absorbed. Only with a complete measurement of the energized recoil electron and the degraded scattered photon can one reconstruct the incident photon’s energy and direction; otherwise, there is an ambiguity corresponding to a conical region. On top of this, it should be possible to detect photons by photoelectric effect in the 100-keV region and below, and in the region above a few tens of MeV, dominated by pair production (see Section \ref{sec:physics}).

For best environmental conditions, a MeV gamma-ray mission  should be launched into a quasi-equatorial (inclination $i < 2.5^\circ$)  LEO at a typical altitude of 550~km. The background environment in such an orbit is now well-known.

To fill the ``MeV gap" and answer many unsolved physics questions, several mission concepts have advanced the state of the art. The designs of   Compton telescopes at sufficient readiness can be
grouped into two families: multi-layer silicon detectors with 2D position resolution at each layer; and thick, large-volume Ge or
CZT detectors with internal 3D position sensitivity. 

\subsection{Multilayer Detectors and the ASTROGAM Concept}

The ASTROGAM concept (see, for example, \cite{eastrogam,scienceea}), proposed in several ESA medium-class calls under the names of ASTROGAM, e-ASTRM, newASTROGAM, matured wide-field Compton/pair techniques with polarimetry (in the US, the AMEGO/COMPAIR class of detectors pursued similar architectures and technologies). 

In the ASTROGAM concept, the instrument combines:
\begin{itemize} [topsep=3pt,parsep=0pt,itemsep=0pt,leftmargin=*,labelsep=6mm,align=parleft]
  \item A silicon (Si) microstrip multiplane tracker to register Compton scatters and pair conversions with fine spatial resolution and fast, low-noise readout.
  \item A pixelated scintillator calorimeter (e.g., CsI(Tl) with SiPM/SDD readout) providing depth-of-interaction and high-efficiency absorption.
  \item Segmented anti-coincidence (AC) panels achieving almost complete charged-particle veto efficiency.
    \item Possibly, a thin, hard X-ray coded mask (effective in an energy range from some {15} to about {40} {keV}) delivering $\sim$arcminute localization of bright transients and persistent sources.
\end{itemize}

Typical dimensions in order to gain one-two orders of magnitude in sensitivity with respect to COMPTEL are about 1 m $\times$ 1 m $\times$ 0.8 m. With such an aspect ratio, the field of view is larger than 2.5 sr above 10 MeV. The typical mass will be of the order of 1 ton.

\begin{figure}[H]
\centering
\includegraphics[width=0.7\linewidth]{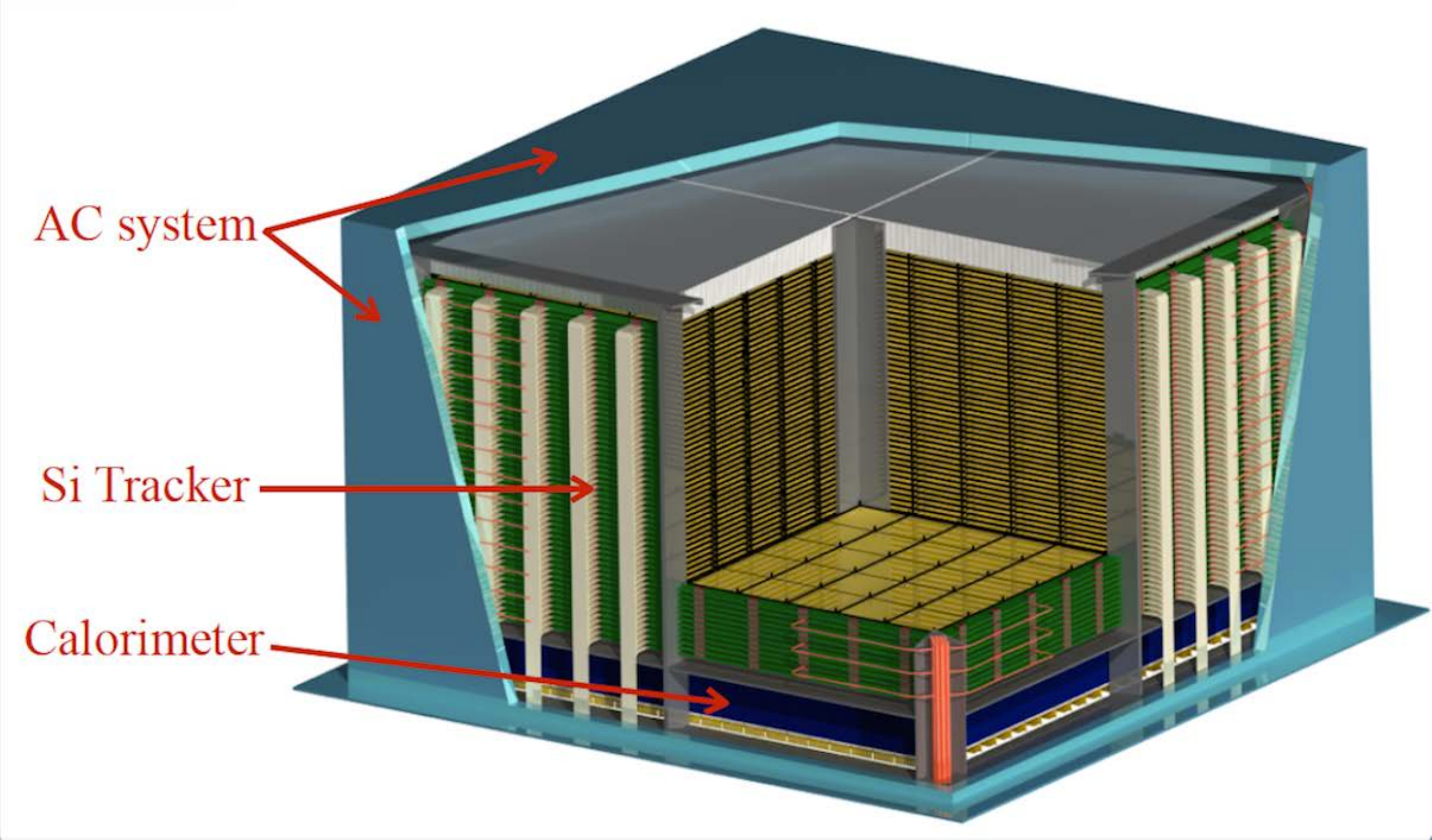}
\caption{Overview of the e-ASTROGAM payload showing the silicon Tracker, the Calorimeter and the Anticoincidence system.}
\label{fig:payload}       
\end{figure}

A realization of the ASTROGAM concept, called e-ASTROGAM, is shown in Figure~\ref{fig:payload}. 

\vspace{-16pt}

\begin{figure}[H]
\centering
\includegraphics[width=0.5\columnwidth]{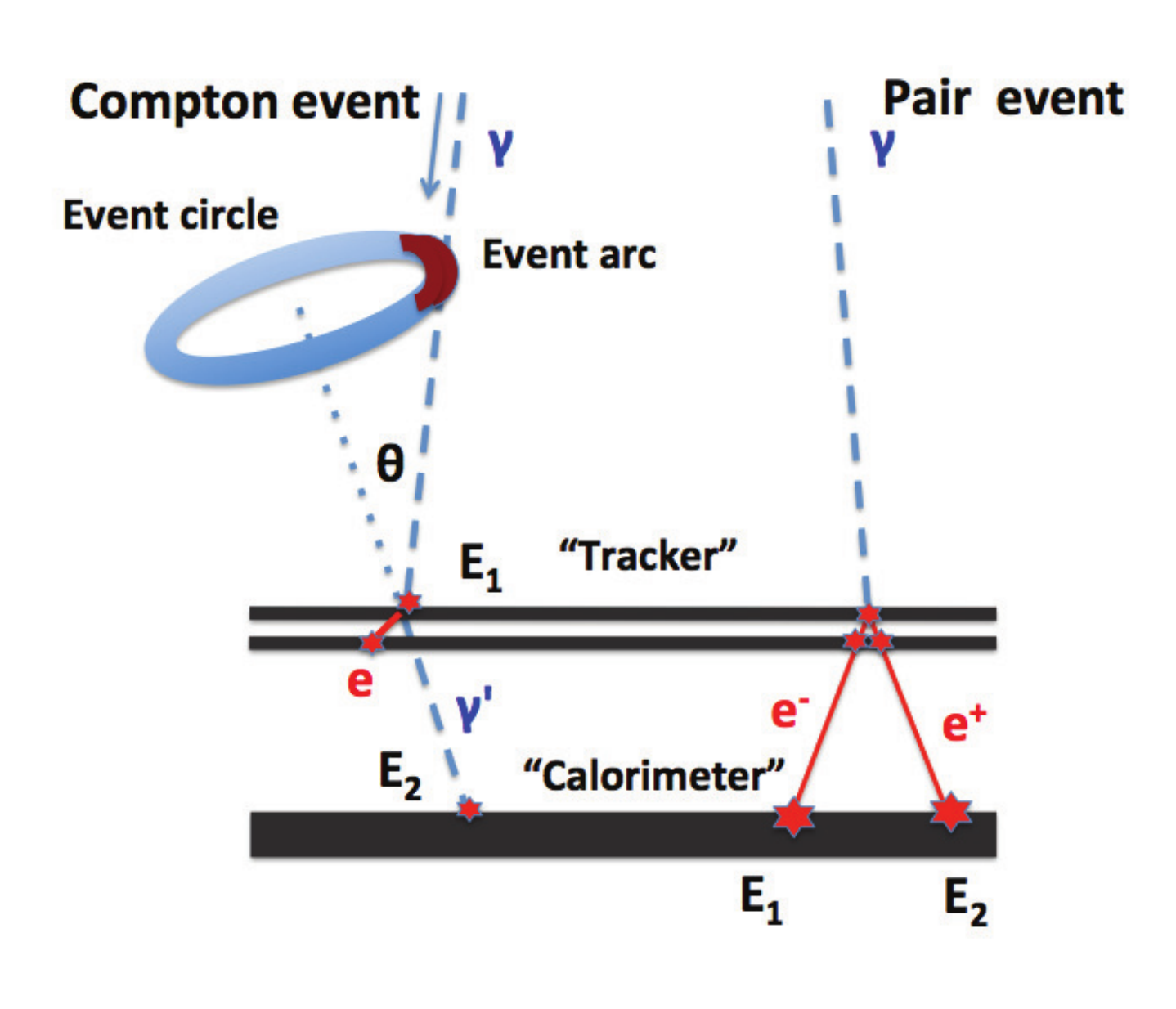}\vspace{-6pt}
\caption{Representative  topologies for a Compton event  ({left}) and  for a pair event ({right}). Photon tracks are shown in pale blue, dashed, and electron and/or positron tracks in red, solid. From {\cite{alexnew}}.}
\label{fig:evt-top}       
\end{figure}

Interactions of photons with matter in the ASTROGAM energy range are dominated by Compton scattering from (below) 0.2~MeV up to about 15 MeV in silicon, and by $e^+e^-$ pair production in the field of a target nucleus at higher energies. ASTROGAM maximizes its efficiency for imaging and spectroscopy of energetic gamma-rays by using both processes. Figure~\ref{fig:evt-top} shows a schematic representation of topologies for Compton and pair events.

For pair-production events, ASTROGAM is similar in design to AGILE and $Fermi$-LAT, but optimized for lower energy. This goal is achieved by eliminating the passive  converters used in both these instruments. This approach reduces gamma-ray conversion efficiency, but it improves the instrument point-spread function (PSF) by reducing absorption and multiple Coulomb scattering of the electron and positron.

The broad point-spread function (PSF) is a primary limiting factor in the science that can be done at energies below 100 MeV by AGILE and $Fermi$-LAT. Pair events produce two main tracks from the created electron and positron. Tracking of the initial opening angle and of the plane spanned by the electron and positron tracks enables direct back-projection of the source position. Multiple scattering of the pair in the tracker material  leads to broadening of the tracks and limits the angular resolution. The nuclear recoil taking up an unmeasured momentum results in an additional small uncertainty. The energy of the gamma-ray is measured using the Calorimeter and information on the electron and positron multiple scattering in the Tracker. Polarization information in the pair domain is given by the azimuthal orientation of the electron-positron plane; in addition to improving the PSF, the use of low-mass tracker planes also enables photon polarization measurements.

%For Compton events, point interactions of the gamma-ray in the Tracker and Calorimeter produce spatially resolved energy deposits, which have to be reconstructed in sequence using the redundant kinematic information from multiple interactions. Once the sequence is established, two sets of information are used for imaging: the total energy and the energy deposit in the first interaction measure the first Compton scatter angle. The combination with the direction of the scattered photon from the vertices of the first and second interactions generates a ring on the sky containing the source direction. Multiple photons from the same source enable a full deconvolution of the image, using probabilistic techniques. For energetic Compton scatters (above $\sim$1 MeV), measurement of the track of the scattered electron becomes possible, resulting in a reduction of the event ring to an arc, hence further improving event reconstruction. Compton scattering angles depend on polarization of the incoming photon, hence careful statistical analysis of the photons for a strong (e.g., transient) source yields a measurement of the degree of polarization of its high-energy emission.

\vspace{-9pt}
\vspace{-16pt}

Detecting gamma rays by Compton scattering  is more complicated than for pair production, because the scattered photon carries a significant amount of the information about the incident photon and thus it needs to be detected too. In practice, a Compton telescope requires two separate photon interactions in order to have a clear detection. The first Compton scattering of the incident photon occurs in one of the Tracker planes, creating an electron and a scattered photon. The Tracker measures the interaction location, the electron energy, and in some cases the electron direction. The scattered photon can be absorbed in the Calorimeter or (with smaller probability) scattered a second time in the Tracker before being absorbed in the Calorimeter where its energy and absorption position are measured; the scattered electron is typically absorbed in the Tracker.

The basic principle of the Compton mode of operation is illustrated in Figure \ref{fig:evt-top}, left. An incident gamma-ray Compton scatters by an angle $\Theta$ in one layer of the Tracker, transferring energy $E_{1}$ to an electron. The scattered photon is then absorbed in the Calorimeter, depositing its energy $E_{2}$, and 
the scattering angle is given by $\cos\Theta = 1- {m_{e}c^{2}}/{E_2}+{m_{e}c^{2}}/{(E_1+E_2)} $, where $m_e$ is the electron mass. 
With this information, one can derive an ``event cone" from which the original photon arrived, defined by the impact point and by an ``event circle''. We will call ``untracked" this sort of Compton events. The uncertainty in the event circle reconstruction is reflected in its width and is due to the uncertainties in direction reconstruction of 
the scattered photon and the energy measurements of the scattered electron $(E_1)$ and the scattered photon $(E_2)$. 
Multiple photons from the same source enable a full deconvolution of the image, using probabilistic techniques. 

For energetic Compton scatters (above $\sim$1 MeV), measurement of the track (and thus of the direction) of the scattered electron becomes possible, resulting in a reduction of the event ring to an arc  with length due to the uncertainty in the electron direction measurement, hence further improving event reconstruction and allowing improved source localization. This kind of event is called  ``tracked", and its direction reconstruction is somewhat similar to that for pair event -- the primary photon direction is reconstructed from the direction and energy of two secondary particles: scattered electron and photon.  Redundant kinematic information from multiple interactions could also help. Compton scattering angles depend on polarization of the incoming photon, hence careful statistical analysis of the photons for a strong (e.g., transient) source yields a measurement of the degree of polarization of its high-energy emission (e.g. \cite{for08}).

Especially for the Compton mode at low energies, but also more broadly over the entire energy range covered by e-ASTROGAM, it is important to keep the amount of passive materials on the top and at the sides of the detector to a minimum, to reduce background  in the field of view and to optimize angular and energy resolutions. In addition, the passive materials between the Tracker layers, and between the Tracker and the Calorimeter, must be minimized for best performance. 

\subsubsection{Silicon Tracker\label{sec:tracker}}

The Si Tracker is the heart of the e-ASTROGAM payload. It is based on the silicon strip detector technology widely employed in medical imaging and particle physics experiments (e.g. ATLAS and CMS at LHC), and already applied to the detection of gamma-rays in space with the AGILE and Fermi missions. The e-ASTROGAM Tracker needs double sided strip detectors (DSSDs) to work also as a Compton telescope.

\subsubsection{Calorimeter}

The e-ASTROGAM Calorimeter is a pixelated detector made of a high-$Z$ scintillation material  for an efficient absorption of Compton scattered gamma-rays and electron-positron pairs.

The simultaneous data set provided by the Silicon Tracker, the Calorimeter and the Anticoincidence system constitutes the basis for the gamma-ray detection. However, thanks to the detector excellent granularity, Calorimeter-only events can be used on board to provide a burst notice and a first approximate localization via fast onboard reconstruction even in the absence of a signal from the Tracker. This data acquisition mode is dedicated to the search for fast transient events such as GRBs and TGFs.

\subsubsection{Angular and Spectral Resolution}

In the pair production domain, the PSF improvement over {\it Fermi}/LAT at energies below a few GeVis due to (i)~the absence of heavy converters in the Tracker, (ii) the light mechanical structure of this kind of detector minimizing the amount of passive material within the detection volume and thus enabling a better tracking of the secondary electrons and positrons, and (iii) the analog readout of the DSSD signals allowing a fine spatial resolution of about 40~$\mu$m ($\sim$1/6 of the microstrip pitch). In the Compton domain, thanks to the fine spatial and spectral resolutions of both the Tracker and the Calorimeter, the ASTROGAM angular resolution will be close to the physical limit induced by the Doppler broadening due to the velocity of the target atomic electrons.

Figure~\ref{fig:Jurgen}  shows an example of the imaging capability of a new MeV--GeV mission in the MeV domain compared to COMPTEL. The e-ASTROGAM synthetic map of the Cygnus region was produced from the third {\it Fermi} LAT (3FGL) catalog of sources detected at photon energies $E_\gamma > 100$~MeV, assuming a simple extrapolation of the measured power-law spectra to lower energies. It is clear from this example that an ASTROGAM-like telescope will substantially eliminate  the confusion issue that severely affected the previous and current generations of gamma-ray telescopes. The ASTROGAM imaging potential will be particularly relevant to study the various high-energy phenomena occurring in the Galactic Center region.

\begin{figure}[H]
%\hspace{1.5cm}
\begin{center}
\begin{minipage}{0.4\linewidth}
\includegraphics[scale=0.35]{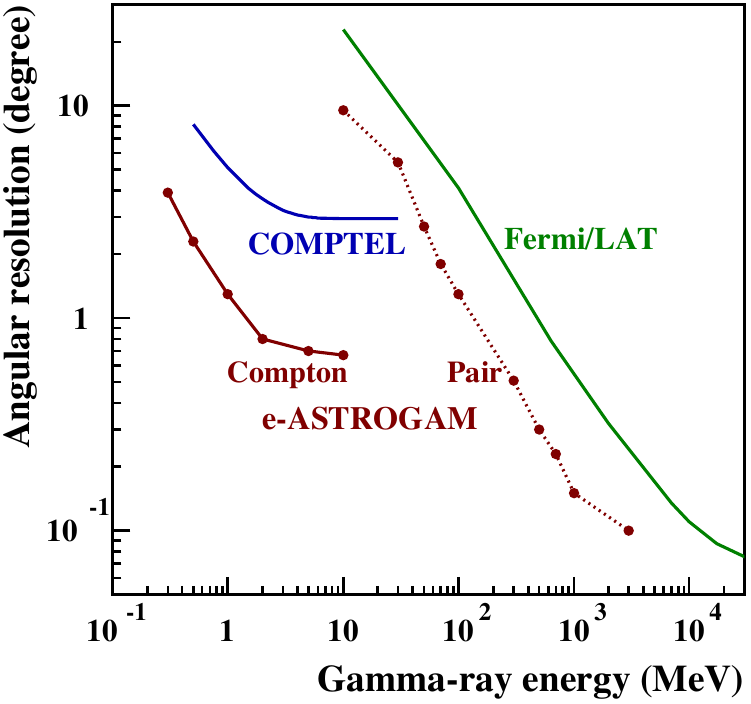}
\end{minipage}
\begin{minipage}{0.4\linewidth}
\includegraphics[scale=0.35]{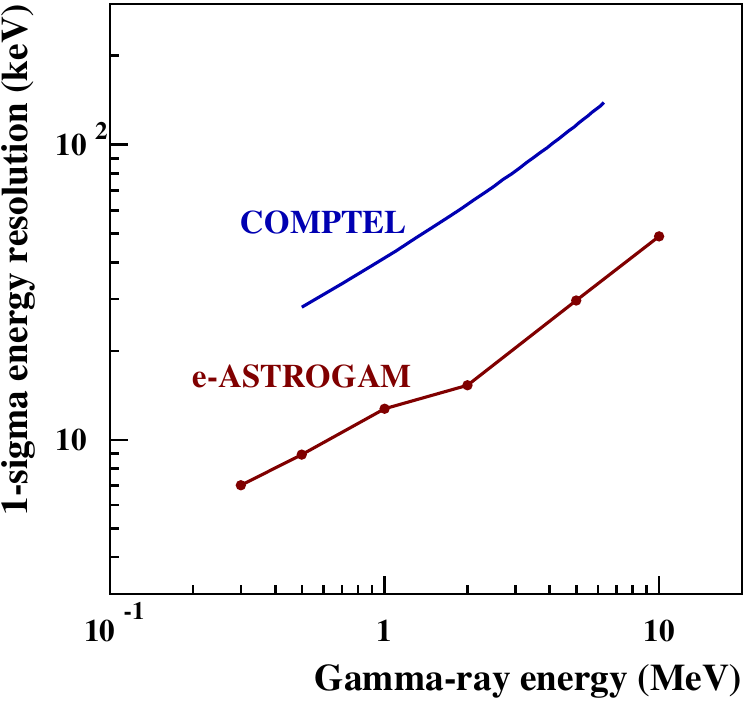}
\end{minipage}
\end{center}
\caption{{({Left panel})---e-ASTROGAM} %Scilight: Please cite the figure in the text and ensure the first citation of each figure appears in numerical order.
 on-axis angular resolution compared to that of COMPTEL and {\it Fermi}/LAT. In the Compton domain, the presented performance of e-ASTROGAM and COMPTEL is the FWHM of the angular resolution measure. In the pair domain, the point spread function (PSF) is the 68\% containment radius for a 30$^\circ$ point source. ({{Right panel}})---1$\sigma$ energy resolution of COMPTEL and e-ASTROGAM in the Compton domain.}
\label{fig:perf}
\end{figure}
\vspace{-28pt}

\subsubsection{Sensitivity}

Improving  sensitivity in the MeV--GeV gamma-ray domain  by one to two orders of magnitude compared to previous missions is the main requirement for the proposed ASTROGAM concept. Such a performance will open an entirely new window for discoveries in the high-energy Universe. 

ASTROGAM can achieve a major gain in sensitivity compared to INTEGRAL/SPI for all gamma-ray lines. An ASTROGAM detector with the size of e-ASTROGAM is expected to reach a line sensitivity for the Doppler-broadened 847 keV line from Type Ia SNe corresponding to an improvement of $\sim 70$ times over SPI. Comparable gains are obtained for other key nucleosynthesis and nuclear lines. This would allow detections of SN 2014J-like events out to $\sim$35 Mpc.
With such an improvement in line sensitivity, ASTROGAM will also (i) provide a much better map of the 511~keV radiation from positron annihilation in the inner Galaxy, (ii) uncover $\sim$10 young, $^{44}$Ti-rich SN remnants in the Galaxy and thus provide new insight on the explosion mechanism of core-collapse SNe (iii)~detect for the first time the expected \cite{CH74} line from $^{22}$Na decay in novae hosted by ONe white dwarfs, (iv)~provide a new constraint on the nuclear equation of state of neutron stars.

Both Compton scattering and pair creation partially preserve the linear polarization information of incident photons. In a Compton telescope, the polarization signature is reflected in the probability distribution of the azimuthal scattering angle. In the pair domain, the polarization information is given by the distribution of azimuthal orientation of the electron-positron plane.

\subsubsection{Technology Readiness}

The ASTROGAM concept is based on the heritage of AGILE and of the Fermi-LAT, with an overall simpler structure. Most components are at or above TRL  6. 

The DSSDs have proven their performance in PAMELA \cite{pamelad} and AMS-02 \cite{ams02d}. Low-power ASICs suitable for the readout of the detectors were already qualified and used \cite{astroh}. 
The low-power SiPM sensors selected for the ACD and the solid-state detectors foreseen for the CsI calorimeter have not yet flown on satellite missions but are in use in suborbital prototypes \cite{ballo} and are planned for use on future space-borne platforms.

The \SI{20}{keV}--\SI{3}{GeV} typical of the ASTROGAM concept bandpass unifies the hard X-ray, MeV, and GeV views with arcminute localizations in hard X-rays, Compton/pair imaging in MeV--GeV, and polarimetry for bright sources and GRBs. Expected improvements over prior missions include order-of-magnitude MeV continuum sensitivity gains, factors of a few to tens in key line sensitivities relative to COSI depending on the transition, and strong constraints from energy-dependent PSF and polarization. This enables:
\begin{itemize}[topsep=3pt,parsep=0pt,itemsep=0pt,leftmargin=*,labelsep=6mm,align=parleft]
  \item Routine detection of short-GRB prompt and early afterglow MeV emission, with polarimetry for the brightest events; fast localizations seed kilonova searches.
  \item Line light curves and profiles in nearby Type~Ia/cc~SNe, novae, and, in favorable cases, kilonovae, constraining nucleosynthesis yields and explosion physics.
  \item Disentangling hadronic vs. leptonic processes in Galactic and AGN jets through MeV SED peaks, polarization, and variability.
  \item Mapping positron annihilation and $\sim$MeV diffuse components in the inner Galaxy with unprecedented fidelity; improved constraints on the Galactic Center excess.
  \item MeV-scale dark-matter and ALP signatures (lines, spectral irregularities) with sensitivity complementary to Fermi and CTAO.
\end{itemize}

\subsection{Crystal Detectors and the COSI Concept}

NASA’s selection of the Compton Spectrometer and Imager (COSI)  for launch in the late 2020s ensures new MeV spectroscopy and Compton imaging with line sensitivity and polarization capability. These steps have de-risked key technologies and clarified requirements for a more ambitious observatory.

COSI is NASA's Astrophysics Small Explorer (SMEX) mission designed to survey the MeV gamma-ray band between $\sim$0.2 and 5~MeV with imaging, spectroscopy, and polarimetry capabilities. Building on a long heritage of balloon flights, the spaceborne COSI will operate in low-Earth orbit with an instantaneous field of view (FoV) exceeding 25\% of the sky, enabling all-sky coverage on daily timescales in survey mode \cite{heasarc_cosi}. The science focus includes mapping positron annihilation at 511~keV in the Milky Way, tracing nucleosynthesis through radioactive lines, measuring polarization in compact objects and gamma-ray bursts (GRBs), and discovering counterparts to multi-messenger sources \cite{tomsick2023,nasa_cosi}.

\begin{figure}[H]
  \centering
  \includegraphics[width=0.85\linewidth]{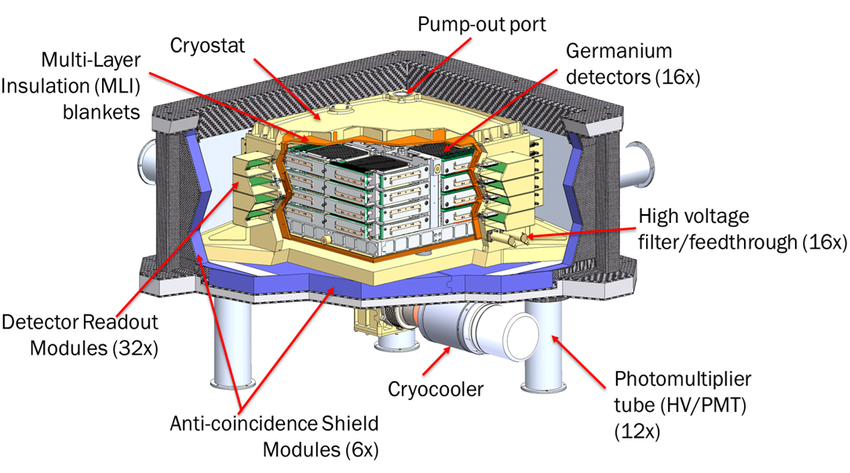}
  \caption{COSI instrument: cutaway view of the detector.}
  \label{fig:cosi}
\end{figure}

COSI is a compact Compton telescope based on an array of 16 high-purity germanium (HPGe) cross-strip detectors (see Figure \ref{fig:cosi}) providing fine 3D interaction localization and excellent spectroscopic performance \cite{cosi_instrument,spie2024_cosi}. Each event is reconstructed via Compton kinematics using the measured positions and energy deposits of multiple interactions, yielding arc-cone imaging on the sky and sensitivity to the azimuthal scattering angle modulation for polarimetry. The detector array is surrounded on the sides and bottom by an active anti-coincidence shield, which suppresses charged-particle and atmospheric backgrounds and enhances the effective FoV for transient detection~\cite{tomsick2023,spie2024_cosi}.

\vspace{-6pt}
\vspace{-16pt}

In the nominal configuration, COSI covers 0.2--5~MeV with a spectral resolution of $\sim$6~keV (FWHM) at 511~keV and $\sim$9~keV at 1.157~MeV, and an angular resolution improving from $\sim$4$^\circ$ at 511~keV to $\sim$2$^\circ$ at 1.809~MeV~\cite{heasarc_cosi}. The instantaneous FoV is $>$25\% of the sky, enabling wide-field monitoring and efficient all-sky surveys \cite{heasarc_cosi}. Polarimetric sensitivity arises naturally from the Compton scattering azimuth distribution, allowing COSI to probe emission geometries and radiation mechanisms in GRBs, magnetars, and black-hole binaries \cite{tomsick2023}.

% Requires \usepackage{graphicx} and \usepackage{hyperref}
% Download the image from:
% https://cosi.ssl.berkeley.edu/wp-content/uploads/2024/12/image-2.png
% and save it next to your .tex file as: cosi_cryostat.png

The detector technology choices imply trade-offs: HPGe offers superior line sensitivity and energy resolution in the classic MeV band, whereas silicon trackers emphasize wide bandpass and effective area at higher energies.

%In the MeV regime, coded-mask instruments excel at pointed line spectroscopy but lack wide FoV; pair telescopes thrive above tens of MeV; silicon hybrids aim for broad coverage; modern HPGe Compton telescopes like COSI offer an excellent blend of high-resolution spectroscopy, wide-field survey capability, and intrinsic polarimetry between 0.2 and 5~MeV. This combination directly targets nucleosynthesis lines, positron astrophysics, and polarization science that have remained sensitivity-limited since the CGRO era.

In the MeV band, coded-mask spectrometers (e.g., INTEGRAL/SPI) deliver keV-class narrow-line resolutions but with a pointed observing mode and limited fully coded FoV ($\sim$$16^\circ$), while pair-conversion telescopes (e.g.,~Fermi-LAT) dominate $\gtrsim$ tens-of-MeV surveys with large FoV ($\sim$$2$--$2.4~\mathrm{sr}$) and a rapidly improving PSF ($\sim$$3.5^\circ$ at $100~\mathrm{MeV}$ $\rightarrow$ $\sim$$0.8^\circ$ at $1~\mathrm{GeV}$ $\rightarrow$ $<$$0.15^\circ$ above $10~\mathrm{GeV}$), and Si Compton/pair hybrids (e-ASTROGAM/AMEGO-class) trade percent-level calorimetric spectroscopy (tens of keV at $\sim$$1~\mathrm{MeV}$) for wide FoV ($\sim$$2.5~\mathrm{sr}$) and broad bandpass, whereas HPGe Compton telescopes such as COSI uniquely combine rather wide-field survey operation (instantaneous FoV $>25\%$ of the sky) with high-resolution line spectroscopy ($\Delta E \sim 6~\mathrm{keV}$ at $511~\mathrm{keV}$; $\sim$$ 9~\mathrm{keV}$ at $1.2~\mathrm{MeV}$), degree-scale Compton imaging ($\sim$$4^\circ$ at $511~\mathrm{keV}$ improving to $\sim$$2^\circ$ at $1.8~\mathrm{MeV}$) and intrinsic polarimetry across $0.2$--$5~\mathrm{MeV}$, directly enabling  nucleosynthesis-lines, positron, and polarization~measurements.

\section{Conclusions}
Gamma-ray astronomy has repeatedly improved our view of the high-energy Universe, from the first Galactic plane detections and pulsars to GeV blazar demography, TeV source populations, and PeV photons. 

Yet a sensitivity gap in the energy region from 100 keV to 3 GeV (the ``MeV gap'') persists as the most underserved electromagnetic window. This gap  limits in particular: (i) line spectroscopy of SNe, novae, and kilonovae; (ii) positron annihilation mapping; (iii) transient polarimetry and afterglow coverage of GRBs and magnetar flares; (iv) dark-matter and ALP searches with distinct MeV phenomenology; and (v) identifications in crowded regions (Galactic Center, Fermi Bubbles). 

The community has advanced multiple concepts to fill this gap. By the time new MeV-GeV missions could hopefully fly after 2030, the landscape will feature
 COSI (a NASA SMEX) delivering Compton imaging/polarimetry in the mid/late 2020s together with  $Fermi$ possibly continuing GeV surveys, all well inserted in a multimessenger~arena.

 COSI's selection ensures a MeV pathfinder, but broader bandpass and deeper sensitivity with arcminute localizations remain compelling.

The programmatic groundwork and the technology readiness now make a wide-band, polarimetric MeV-GeV mission with autonomous localization capabilities like the ASTROGAM concept realistic to close the MeV gap and to act as a fundamental actor for multi-messenger astrophysics in the 2030s and beyond, while also delivering autonomously transformative high-energy science.

 \section*{Acknowledgements}
I would like to acknowledge the entire MeV--GeV gamma-ray community, which for many years has worked to advance the scientific and technological readiness of the field and its instrumentation. I warmly thank in particular my colleagues and friends who have contributed to refining and strengthening the ASTROGAM concept. These efforts have  shaped and supported this article.

	\newcommand{\etal}{\textit{et al. }}
	
	\small
%	\bibliographystyle{scilight}
	%=====================================
	% References, variant A: internal bibliography
	%=====================================

\end{document}